\documentclass[aps,prd,preprint,groupedaddress,showpacs]{revtex4-1}
\usepackage{amssymb,amsmath,graphicx,ulem}
\usepackage{verbatim,color}

\begin{document}
\title{Subvacuum effects of the quantum field on the dynamics of a test particle}
\author{Tai-Hung Wu}
\author{Jen-Tsung Hsiang}

\email{cosmology@gmail.com}
\author{Da-Shin Lee}
\email{dslee@mail.ndhu.edu.tw} \affiliation{Department of Physics,
National Dong Hwa University, Hualien, Taiwan, R.O.C.}

\begin{abstract}
We study the effects of the electromagnetic subvacuum fluctuations on the dynamics of a nonrelativistic charged particle in a wavepacket. The influence from the quantum field is expected to give an additional effect to the velocity uncertainty of the particle. In the case of a static wavepacket, the observed velocity dispersion is smaller in the electromagnetic squeezed vacuum background than in the normal vacuum background. This leads to the subvacuum effect. The extent of reduction in velocity dispersion associated with this subvacuum effect is further studied by introducing a switching function. It is shown that the slow switching process may make this subvacuum effect insignificant. We also point out that when the center of the wavepacket undergoes non-inertial motion, reduction in the velocity dispersion becomes less effective with its evolution, no matter how we manipulate the nonstationary quantum noise via the choice of the squeeze parameters. The role of the underlying fluctuation-dissipation relation is discussed.
\end{abstract}

\maketitle

\allowdisplaybreaks

\section{Introduction}
Manipulating the quantum field may give rise to suppression of its
vacuum fluctuations, leading to a subvacuum phenomenon. One of the
known examples is the existence of negative energy density. It has
been shown~\cite{EP} that the renormalized expectation value of the
energy density operator can become negative in some spacetime
region.  This negative energy density may imply exotic phenomena
such as the traversable wormholes~\cite{MO} and the warp
drive~\cite{AL}. It is also known~\cite{FOR,FO1,PF,FE} that the
renormalized local energy density/flux can not be arbitrarily
negative for an arbitrarily long period of time. Similar to the
uncertainty principle, there exists an inequality, constraining
negativeness and duration that might unruledly violate the second
law of thermodynamics~\cite{FO,DA} and the cosmic censorship
hypothesis by creating naked singularities~\cite{FOR}. The squeezed
vacuum state of the electromagnetic field is an example that may
cause subvacuum phenomena.

In the laboratory, squeezed light is produced by a nonlinear-optics technique of ``squeezing'', and then a set of rapidly rotating mirrors further separate the positive and negative energy pulses from each other, so that the negative energy density can be possibly observed~\cite{DAP,DA}. In the early universe the squeezed vacuum state of the quantum matter field may be evolved from its initial vacuum state by amplification of the vacuum fluctuations through the processes of particle creation, for example, during an inflation epoch~\cite{BO}.

Detection of the subvacuum fluctuations of the quantum scalar field has been studied by considering the response of a static particle detector, whose monopole moment couples to the field~\cite{DAO}. Various switching functions are introduced, but only one single time scale is used to characterize the processes of switching on/off and measuring. This switching process results in excitations of the detector via interaction with the quantum field. In particular, the coupling to the squeezed vacuum state of the field may suppress the rate of excitations of the detector to a level less than what would be caused by a normal vacuum state. The quantum inequality associated with this subvacuum effect is then discussed. Some other subvacuum phenomena have been proposed in laboratory experiments~\cite{MA,FOG,HF}.

Here we would like to study the subvacuum effects of the quantized electromagnetic field, so a natural choice of the detector is a charged particle since it has a well-defined particle-field interaction. More specifically, we wish to explore the effects of electromagnetic squeezed vacuum on the dynamics of the charged particle, which is prepared in a wavepacket. The charged particle is considered as the system of interest, and the degrees of freedom of the fields as the environment. The linear coupling with the electromagnetic gauge potential allows us to integrate out the field variables exactly. Within the context of the closed-time-path formalism~\cite{FV,SK,GR1,WU,JO,HUP,JT,HLW,HWL,BA}, we will obtain the influence functional that encodes all effects from the fields upon the particle.

The influence of the electromagnetic field fluctuations is expected to give an additional effect to the dynamics of the particle. Since the charged particle is never observed without being affected by the electromagnetic zero-point fluctuations, the observed velocity dispersion should already include the additional effect due to normal vacuum fluctuations. Thus we define the renormalized velocity dispersion of the free particle by absorbing the normal vacuum contributions into the intrinsic velocity dispersion due to the finite wavepacket size. Then, we investigate how this renormalized velocity dispersion can be possibly reduced by the squeezed vacuum fluctuations so that it is smaller than its counterpart in normal vacuum fluctuations of the fields. This leads to the subvacuum phenomenon. Our approach of treating a nonrelativistic charged particle quantum-mechanically should give leading-order results when the cutoff energy scale of radiation fields is consistently set at the inverse of the width of the wavepacket, much smaller than the rest mass energy of the particle. It is no doubt that a more precise quantitative evaluation certainly requires the full QED study. This, in spirit, follows a similar treatment of the Lamb shift, proposed by Weldon in a more heuristic way~\cite{ITZ}, as well as by Bethe in terms of the time-dependent perturbation theory~\cite{SAK}. There the associated energy shift of hydrogen states can be understood mainly arising from the influence of the quantum fluctuations of electromagnetic fields on nonrelativistic quantum-mechanical, bound electrons. Their approach turns out to yield an estimated energy shift with the correct order of magnitude as compared with the results from the full QED calculations~\cite{ITZ}.

Next, for a more detailed analysis of the subvacuum phenomena, we incorporate the switching function, as a consequence of the finite-time switching-on/off process of interaction between the charged particle and electromagnetic squeezed modes. We first consider a sudden switching-on process, and then generalize the switching process to the case with a finite switching time by introducing a suitable switching function. We can derive an inequality associated with this subvacuum phenomenon. Finally, we consider that the center of a wavepacket undergoes non-inertial motion where the electromagnetic self-force will give rise to a weak damping effect on the evolution of the trajectory. We find that non-trivial motion tends to modify the result of the velocity dispersion in such a way that it becomes less effective to suppress the environmental noise than the case of inertial motion.

Our presentation is organized as follows. In Sec.~\ref{sec1}, we briefly introduce the closed-time-path formalism in order to describe the evolution of the reduced density matrix of a nonrelativistic charged particle coupling to the squeezed vacuum state of the quantized electromagnetic field. The field variables are traced out to obtain the influence functional. The reduced density matrix of the particle is formally derived under the WKB approximation, in particular, when the particle is in non-inertial motion. The technique of solving the Langevin equation in the dipole approximation has been introduced in Ref.~\cite{HWL}. We then consider the velocity dispersion of the charged particle in Sec.~\ref{sec2}, where its full-time evolution is studied both analytically and numerically. The results are summarized and the implications are discussed in Sec.~\ref{sec3}.

The Lorentz-Heaviside units and the convention $\hbar=c=1$ will be used unless otherwise mentioned. The signature of the metric is $\operatorname{diag}\{\eta_{\mu\nu}\}=(+1,-1,-1,-1)$.

\section{reduced density matrix}\label{sec1}
When a nonrelativistic  particle with charge $e$ and mass $m_0$
interacts with the electromagnetic field, its dynamics is described
by the Lagrangian
\begin{align}\label{E:lagrangian-charge}
    L[\mathbf{q},\mathbf{A}_{\mathrm{T}}]=\frac{1}{2}m_0 \dot{\mathbf{q}}^2-V(\mathbf{q})-&\frac{1}{2}
    \int\!d^3\mathbf{x}\,d^3\mathbf{y}\;\varrho(x;\mathbf{q})G(\mathbf{x},\mathbf{y})\varrho(y;\mathbf{q})\notag\\
    &\qquad\qquad\qquad+\int\!d^3\mathbf{x}\;\left[\frac{1}{2}(\partial_\mu\mathbf{A}_{\mathrm{T}})^2+\mathbf{j}\cdot\mathbf{A}_{\mathrm{T}}\right]\,,
\end{align}
where the Coulomb gauge $\mathbf{\nabla}\cdot\mathbf{A}_{\mathrm{T}}=0$ is chosen. $\mathbf{A}_{\mathrm{T}}$ and $\mathbf{q}$ are the transverse components of the gauge potential, and the position of the charge, respectively. The instantaneous Coulomb Green's function $G(\mathbf{x},\mathbf{y})$ satisfies the Gauss law. The charge and current densities take the forms, respectively,
\begin{equation*}
     \varrho(x;\mathbf{q}(t))=e\,\delta^{(3)}(\mathbf{x}-\mathbf{q}(t))\,,\qquad\qquad\mathbf{j}(x;\mathbf{q}(t))=e\,\dot{\mathbf{q}}(t)\,\delta^{(3)}(\mathbf{x}-\mathbf{q}(t))\,.
\end{equation*}
The density matrix of the combined charge-field system, $\hat{\rho}(t)$, evolves unitarily according to the functional Liouville equation. The effects of the field on the particle's dynamics can be realized by the reduced density matrix $\hat{\rho}_r$ of the particle, which is obtained by tracing out the field variables in the total density matrix $\hat{\rho}(t)$. When the initial time $t=t_i$ is in the remote past ($t_i \rightarrow -\infty$), the system and environmental fields are conveniently assumed to be uncorrelated,
\begin{equation*}
    \hat{\rho}(t_i)=\hat{\rho}_{e}
    (t_i)\otimes\hat{\rho}_{\mathbf{A}_{\mathrm{T}}}(t_i)\,,
\end{equation*}
where the field is originally in the vacuum state. This vacuum state will be further modified later at the time $t_0$ by the process of ``squeezing", resulting in the squeezed vacuum state.

Under this assumption of the factorizable initial condition, the reduced density matrix at later time $t_{f}$ takes a simple form,
\begin{align}
     \rho_r(\mathbf{q}_f,\tilde{\mathbf{q}}_f,t_f)&=\int\!d\mathbf{A}_{\mathrm{T}}\;\bigl<\mathbf{q}_f,\mathbf{A}_{\mathrm{T}}\big|\rho(t_f)\big|\tilde{\mathbf{q}}_f,\mathbf{A}_{\mathrm{T}}\bigr>\notag\\
                   &=\int\!d^3\mathbf{q}_1\,d^3\mathbf{q}_2\;\mathcal{J}(\mathbf{q}_f,\tilde{\mathbf{q}}_f,t_f;\mathbf{q}_1,\mathbf{q}_2,t_i)\,\rho_{e}(\mathbf{q}_1,\mathbf{q}_2,t_i)\,,\label{E:ddfff}
\end{align}
where the propagating function $\mathcal{J}(\mathbf{q}_f,\tilde{\mathbf{q}}_f,t_f;\mathbf{q}_1,\mathbf{q}_2,t_i)$ is defined as
\begin{equation}\label{E:dfdfdfdf}
    \mathcal{J}(\mathbf{q}_f,\tilde{\mathbf{q}}_f,t_f;\mathbf{q}_1,\mathbf{q}_2,t_i)=\int^{\mathbf{q}_f}_{\mathbf{q}_1}\!\!\mathcal{D}\mathbf{q}^+\!\!\int^{\tilde{\mathbf{q}}_f}_{\mathbf{q}_2}\!\!\mathcal{D}\mathbf{q}^-\;\exp\left[i\int_{t_i}^{t_f}dt\left(L_{e}[\mathbf{q}^+]-L_{e}[\mathbf{q}^-]\right)\right]\mathcal{F}[\mathbf{j}^+_{\mathrm{T}},\mathbf{j}^-_{\mathrm{T}}]\,,
\end{equation}
and the Lagrangian $L_e[\mathbf{q}]$ takes the form
\begin{equation}\label{lagrangian-e}
    L_e\bigl[\mathbf{q}\bigr]=\frac{1}{2}m_0 \dot{\mathbf{q}}^2-V(\mathbf{q})-\frac{1}{2}\int\!d^3\mathbf{x}\,d^3\mathbf{y}\;\varrho(x;\mathbf{q})\,G(\mathbf{x},\mathbf{y})\,\varrho(y;\mathbf{q})\,.
\end{equation}
The influence functional, $\mathcal{F}[\mathbf{j}^+_{\mathrm{T}},\mathbf{j}^-_{\mathrm{T}}]$, which contains full information about the effects of the environmental fields on the particle, is given by
\begin{equation*}
    \mathcal{F}[\mathbf{j}^+_{\mathrm{T}},\mathbf{j}^-_{\mathrm{T}}]=\operatorname{Tr}_{\mathbf{A}_{\mathrm{T}}}\left\{ U(t_f,t_i\,;\mathbf{j}^+_{\mathrm{T}})\,\hat{\rho}_{\mathbf{A}_{\mathrm{T}}}(t_i)\,U^{-1} (t_f,t_i \,;\mathbf{j}^-_{\mathrm{T}}) \right\}\,,
\end{equation*}
where $U(t_f,t_i\,;\mathbf{j}_{\mathrm{T}})$ is an evolution operator of the free $\mathbf{A}_{\mathrm{T}}$ field driven by a classical current density $\mathbf{j}_{\mathrm{T}}$. Since the charge linearly couples to the electromagnetic field, the field variables can be integrated out exactly. The laborious derivations of the influence functional can be found in Ref.~\cite{HUP}, and the resulting expression of $\mathcal{F}\left[\mathbf{j}^{+},\mathbf{j}^{-}\right]$ is then obtained in terms of real-time Green's functions of the vector potentials, explicitly given in Appendix~\ref{appena}.

The squeezed vacuum states can be constructed out of the normal vacuum state by application of the squeeze operator $\mathcal{S}( \zeta_{\mathbf{k}})$,
\begin{equation*}
 \left|0\right>_{\zeta_{\mathbf{k}}}  =\mathcal{S}( \zeta_{\mathbf{k}})\left|0\right> \,.
\end{equation*}
The squeeze operator $\mathcal{S}( \zeta_{\mathbf{k}})$ is defined by
\begin{equation}
    \mathcal{S}(\zeta_{\mathbf{k}}) =\exp\bigg[\frac{\zeta^*_{\mathbf{k}}}{2}a^2_{\lambda\,\mathbf{k}}-\frac{\zeta_{\mathbf{k}}}{2} a^{\dagger\,2}_{\lambda\,\mathbf{k}}\bigg]\,,\qquad\text{and}\qquad\zeta_{\mathbf{k}}=r_{\mathbf{k}}\,e^{i\theta_{\mathbf{k}}} \,,
\end{equation}
and the wave vector $\mathbf{k}$ labels the modes of the squeezed vacuum states. The squeeze parameter $\zeta_{\mathbf{k}}=r_{\mathbf{k}}\,e^{i\theta_{\mathbf{k}}}$ is an arbitrary complex number with $r_{\mathbf{k}}\geq0$ and $\theta_{\mathbf{k}}\in\mathfrak{R}$. The creation and annihilation operators satisfy the canonical commutation relations
\begin{equation*}
     \left[a_{\lambda\,\mathbf{k}},a^{\dagger}_{\lambda'\,\mathbf{k}'}\right]=\delta_{\lambda\lambda'}\delta^{(3)}(\mathbf{k}-\mathbf{k}')\,,
\end{equation*}
and the operator $a_{\lambda\,\mathbf{k}}$ annihilates the normal vacuum $\lvert 0\rangle$, that is, $a_{\lambda\,\mathbf{k}}\lvert0\rangle=0$. The plane-wave expansion of the vector potential
is of the form
\begin{equation}\label{mode}
     \mathbf{A}_{\mathrm{T}}(x)=\int\!\frac{d^3\mathbf{k}}{(2\pi)^{3/2}}\,\frac{1}{\sqrt{2\omega}}\sum_{\lambda=1,2}\hat{\boldsymbol{\epsilon}}_{\lambda\,\mathbf{k}}\, a_{\lambda\,\mathbf{k}}\,e^{i\mathbf{k}\cdot\mathbf{x}-i\omega t}+\text{h.c.}\,,
\end{equation}
with $\omega=\left|\mathbf{k}\right|$, and the polarization unit vectors $\hat{\boldsymbol{\epsilon}}_{\lambda\,\mathbf{k}}$ obey the transversality condition,
\begin{equation*}
     \sum_{\lambda=1,2}\hat{\boldsymbol{\epsilon}}^{i}_{\lambda\,\mathbf{k}}\,\hat{\boldsymbol{\epsilon}}^{j}_{\lambda\,\mathbf{k}}=\delta^{ij}-\frac{k^ik^j}{\left|\mathbf{k}\right|^2}\,.
\end{equation*}
With the help of the Baker-Campbell-Hausdorff formula, we readily find the unitary transformations of the creation and annihilation operators due to the squeeze operator
$\mathcal{S}\left(\zeta_{\mathbf{k}}\right)$,
\begin{equation*}
     \mathcal{S}^{\dagger}\left(\zeta_{\mathbf{k}}\right)a_{\lambda\,\mathbf{k}}\,\mathcal{S}\left(\zeta_{\mathbf{k}}\right)=\mu_{\mathbf{k}}a_{\lambda\,\mathbf{k}}-\nu_{\mathbf{k}}a^{\dagger}_{\lambda\,\mathbf{k}}\,,\qquad\text{and}\qquad\mathcal{S}^{\dagger}\left(\zeta_{\mathbf{k}}\right)a^{\dagger}_{\lambda\,\mathbf{k}}\mathcal{S}\left(\zeta_{\mathbf{k}}\right)=\mu_{\mathbf{k}}a^{\dagger}_{\lambda\,\mathbf{k}}-\nu^*_{\mathbf{k}}a_{\lambda\,\mathbf{k}}\,,
\end{equation*}
with $\mu_{\mathbf{k}}=\cosh r_{\mathbf{k}}$, $\nu_{\mathbf{k}}=\sinh r_{\mathbf{k}}\,e^{i\theta_{\mathbf{k}}}$, $\eta_{\mathbf{k}}=\left|\nu_{\mathbf{k}}\right|$, and $\mu^2_{\mathbf{k}}-\left|\nu_{\mathbf{k}}\right|^2=\mu^2_{\mathbf{k}}-\eta_{\mathbf{k}}^{2}=1$. The expectation values of the creation and the annihilation operators in the squeezed vacuum state are respectively given by
\begin{equation*}
    \left<0\right|\mathcal{S}^{\dagger}\left(\zeta_{\mathbf{k}}\right)a_{\lambda\,\mathbf{k}}\,\mathcal{S}\left(\zeta_{\mathbf{k}}\right)\left|0\right>=0\,,\qquad\qquad\left<0\right|\mathcal{S}^{\dagger}\left(\zeta_{\mathbf{k}}\right)a^{\dagger}_{\lambda\,\mathbf{k}}\,\mathcal{S}\left(\zeta_{\mathbf{k}}\right)\left|0\right>=0\,.
\end{equation*}
Moreover, we have
\begin{align}\label{E:number}
     &\left<0\right|\mathcal{S}^{\dagger}\left(\zeta_{\mathbf{k}}\right)a^2_{\lambda\,\mathbf{k}}\,\mathcal{S}\left(\zeta_{\mathbf{k}}\right)\left|0\right>=-\mu_{\mathbf{k}}\nu_{\mathbf{k}}\,,\qquad\qquad\left<0\right|S^{\dagger}\left(\zeta_{\mathbf{k}}\right)a^{\dagger\,2}_{\lambda\, \mathbf{k}}\,\mathcal{S}\left(\zeta_{\mathbf{k}}\right)\left|0\right> =-\mu_{\mathbf{k}}\nu^*_{\mathbf{k}}\,,\notag\\
     &\left<0\right|\mathcal{S}^{\dagger}\left(\zeta_{\mathbf{k}}\right)a^{\dagger}_{\lambda\,\mathbf{k}}a^{\vphantom{\dagger}}_{\lambda\,\mathbf{k}}\,\mathcal{S}\left(\zeta_{\mathbf{k}}\right)\left|0\right>=\eta_{\mathbf{k}}^2\,.
\end{align}
From Eq.~\eqref{E:number}, we see that the squeezed vacuum state is different from a normal vacuum state, and it contains $\eta_{\mathbf{k}}^2$ photons on average for each mode $\mathbf{k}$. Accordingly, all Green's functions can be evaluated from the expressions of these expectation values.

It is quite straightforward to cast the reduced density \eqref{E:ddfff} into the form,
\begin{equation}
     \rho_{r}(\mathbf{q}_{f},\tilde{\mathbf{q}}_{f},t_{f})=\int\mathcal{D}\xi\;\mathcal{P}[\xi]\,\rho_{r}(\mathbf{q}_{f},\tilde{\mathbf{q}}_{f},t_{f};\xi)\,,
\label{Srho}
\end{equation}
in which $\rho_{r}(\mathbf{q}_{f},\tilde{\mathbf{q}}_{f},t_{f};\xi)$ is a reduced density of the particle under the influence of some realization of the environmental stochastic noise $\xi$, and
\begin{equation} \rho_{r}(\mathbf{q}_{f},\tilde{\mathbf{q}}_{f},t_{f};\xi)=\int_{-\infty}^{\infty}d^3\mathbf{q}_{1}\,d^3\mathbf{q}_{2}\int^{q_{f}}_{q_{1}}\mathcal{D}\mathbf{q}^{+}\!\!\int_{q_{2}}^{\tilde{q}_{f}}\mathcal{D}\mathbf{q}^{-}\;e^{i\,S_{\xi}[\mathbf{q}^{+},\mathbf{q}^{-};\xi]}\rho_{e}(\mathbf{q}_{1},\mathbf{q}_{2},t_{i})\,.
\end{equation}
Here $S_{\xi}$ is the stochastic coarse-grained effective action defined by
\begin{equation}
    S_{\mathbf{\xi}}[\mathbf{q}^{+},\mathbf{q}^{-}
    ;\xi]=\operatorname{Re}\left\{S_{\text{CG}}[\mathbf{q}^{+},\mathbf{q}^{-}]\right\}-e\int_{t_{i}}^{t_{f}}dt\;(
    \mathbf{q}^{+}-\mathbf{q}^{-})^{k}\left(\delta^{kl}\frac{d}{dt}-q^{l}\nabla^{k}\right)\xi^{l}
    \,  \label{s:xi}
\end{equation}
with the coarse-grained effective action $S_{\text{CG}}$ constructed as follows
\begin{equation}
    S_{\text{CG}}[\mathbf{q}^{+},\mathbf{q}^{-}]=
S_{e}[\mathbf{q}^{+}]-S_{e}[\mathbf{q}^{-}]-i \ln
\mathcal{F}[\mathbf{j}^+_{\mathrm{T}},\mathbf{j}^-_{\mathrm{T}}] \,,
\label{s:cg}
\end{equation}
in which $S_{e}[\mathbf{q}]$ is the action corresponding to the Lagrangian \eqref{lagrangian-e}. More detailed derivations can be found in Appendix~\ref{appena}. The average of a quantum operator associated with the charged particle is thus defined by
\begin{equation}
    \langle {\cal{O}}\rangle =\int\mathcal{D}\xi\;\mathcal{P}[\xi]\int_{-\infty}^{\infty} d^3
    \mathbf{q}_{f}\;{\cal{O}} \,
    \rho_{r}(\mathbf{q}_{f},\mathbf{q}_{f},t_{f};\xi)\,,
    \label{O}
\end{equation}
with the probability distribution functional $\mathcal{P}[\xi(t)]$ given by
\begin{equation}
    \mathcal{P}[\xi(t)]=\exp\left\{-\frac{\hbar}{2}\int_{t_i}^{t_f}dt\int_{t_i}^{t_f}dt'\;\left[\xi^i(t)\,G_{H}^{ij}{}^{-1}\left[{\bf q}(t),{\bf q}(t');t,
    t'\right]\,\xi^j(t')\right]\right\}\, .
\label{noisedistri}
\end{equation}
The Hadamard function $G_{H}^{ij}$ will be defined later in \eqref{Hadamrd}.

If we consider the charged particle is initially prepared in a Gaussian wavepacket, then in the case of the static motion the remaining integrals are all Gaussian so that the reduced density matrix can be obtained exactly. However, when the center of the wavepacket undergoes non-inertial motion, we have to formally use the WKB approximation to evaluate the path integral expression in $\rho_{r}(\mathbf{q}_{f},\mathbf{q}_{f},t_{f};\xi)$ about the particle's classical trajectory, which is determined by the Langevin equation~\cite{HWL},
\begin{equation}\label{lange}
    m_0 \ddot{q}^i+\nabla^iV(\mathbf{q}(t))=e \big[ E^i (\mathbf{q}) +\epsilon_{ijk} \dot{q}^j(t) B^k (\mathbf{q})\big]-\hbar\, e \, \left( \delta^{il} \frac{d}{d t} - \dot{q}^l
(t)\nabla^i \right) \, \xi^l (t)\,. \notag
\end{equation}
The electromagnetic fields can be expressed in
terms of the Lienard-Wiechert potential $\mathbf{A}_{\rm LW} $ in the Coulomb gauge
due to motion of the charge,
\begin{equation}
A^i_{\rm LW} (\mathbf{q}) = - e \int_{t_i}^{t_f }dt'\;
G_{R}^{lj} \left[\mathbf{q}(t),\mathbf{q}(t');
t-t'\right]\,\dot{q}^j (t') \, ,
\end{equation}
so they are given by
\begin{equation}
\mathbf{E}=-\frac{\partial}{\partial t} \mathbf{A}_{\rm LW} \, , \qquad
\mathbf{B}=\nabla_{\mathbf{q}} \times \mathbf{A}_{\rm LW} \, .
\end{equation}
The stochastic noise $\xi^{i}(t)$ satisfies the statistical
correlation,
\begin{equation}\label{noisec}
    \langle\xi^i(t)\rangle =0\,,\qquad\qquad\langle\xi^i (t)\xi^j(t')\rangle=\frac{1}{\hbar} G_{H}^{ij} \left[\mathbf{q}(t),\mathbf{q}(t');t,t'\right]\,,
\end{equation}
and the higher moments of $\xi^{i}(t)$ vanishes. This semiclassical approximation can be justified as long as the initial width of the wavepacket is much larger than the de Broglie wavelength of the particle, that is, $\sigma_0 \gg \lambda_{\rm dB}$, where the uncertainty of the velocity is then much smaller than its mean value. Here the trajectory, the center of the prepared wavepacket, is stochastic in nature owing to the quantum fluctuations of the field. It is in analogy to the Brownian motion~\cite{YF}. This stochastic approach has consistently incorporated both fluctuation and dissipation backreaction of the environmental fields on the particle.

This Langevin equation in Eqs.\eqref{lange} and \eqref{noisec}
encompasses the effects of fluctuation and dissipation backreaction
on the motion of the center of the wavepacket from the quantized
electromagnetic fields via the kernels $G_{H}^{ij}$ and
$G_{R}^{ij}$, which are respectively defined by 
\begin{align}
   &  \hbar\,G_{H}^{ij}(\mathbf{q}(t),\mathbf{q}(t'); t,t')=\frac{1}{2}\,\bigl<\left\{A_{\mathrm{T}}^i(\mathbf{q}(t),t),A_{\mathrm{T}}^j(\mathbf{q}(t'),t')\right\}\bigr>\label{Hadamrd}\\
        &\quad\quad =\frac{1}{2}\int\!\frac{d^3 \mathbf{k}}{(2 \pi )^3}\;\frac{1}{2\omega}\biggl(\delta^{ij}-\frac{k^{i}k^{j}}{\omega^2}\biggr)\biggl\{\Bigl[2\eta^2_{\mathbf{k}}\, \theta(t-t_0) \, \theta(t'-t_0) +1 \Bigr]\,e^{-i\omega(t-t')}e^{i\mathbf{k}\cdot[\mathbf{q}(t)-\mathbf{q}(t')]}\biggr.\notag\\
        &\qquad\qquad\qquad\qquad\qquad\qquad\biggl.-\,2\eta_{\mathbf{k}}\mu_{\mathbf{k}}\,\theta(t-t_0) \, \theta (t'-t_0) \, e^{i\theta_{\mathbf{k}}-i\omega(t+t')}e^{i\mathbf{k}\cdot[\mathbf{q}(t)+\mathbf{q}(t')]}\biggr\}+\text{c.c.}\notag\\
        &\quad\quad=\hbar\,G_{H,\,st}^{ij}(x,x')+\hbar\,G_{H,\,ns}^{ij}(x,x')\,,\notag\\
 &   \hbar\,G_{R}^{ij}(\mathbf{q}(t),\mathbf{q}(t');t,t')=i\,\theta(t-t')\,\bigl<\left[A_{\mathrm{T}}^i(\mathbf{q}(t),t),A_{\mathrm{T}}^j(\mathbf{q}(t'),t')\right]\bigr>\label{commutator} \\
        &\quad\quad=i\,\theta (t-t')\int\!\frac{d^3\mathbf{k}}{(2\pi)^3}\;\frac{1}{2\omega}\biggl(\delta^{ij}-\frac{k^{i}k^{j}}{\omega^2}\biggr)\biggl\{e^{-i\omega(t-t')}e^{i\mathbf{k}\cdot[\mathbf{q}(t)-\mathbf{q}(t')]}\biggr\}+\text{c.c.}\,,\notag
\end{align}
in which the squeezed vacuum modes are turned on at $t=t_0$. Apparently, there are two distinct contributions to the Hadamard function $G_{H}^{ij}$ of the squeezed vacuum state. The first term $G_{H,\,\text{st}}^{ij}$ in the curly brackets comes from the stationary component of the noise. However, there exists a nonstationary component $G_{H,\,\text{ns}}^{ij}$ due to the fact that the squeezed state is generated by nonlinear processes of particle creation in which the time-translational invariance is broken. In contrast, the retarded Green's function, accounting for the dissipation backreaction, is independent of the state of the environmental fields in the case of linear coupling, and will contributes to the known self-force of the charged particle, given by a third-order time derivative of its position~\cite{HWL}. It will be shown that nonstationary noise may result in the transient effects on the particle's dynamics over the time scales determined by the frequency bandwidth of the squeezed modes. On the other hand, since the dissipation backreaction is significant when the dynamics of the charge evolves into the relaxation regime, the stationary part of the Hadamard function may become important to dynamically stabilize this evolution, and thus we may derive a fluctuation-dissipation relation that links the retarded Green's function to the stationary component of the Hadamard function,
\begin{equation}\label{E:df-vac}
    G_{H,\,\text{st}}^{ij}[\mathbf{q}(t),\mathbf{q}(t');\omega]
    =(2\eta_{\mathbf{k}}^2 +1)\left[\theta(\omega)-\theta(-\omega)\right]\,\operatorname{Im}G_{R}^{ij}[\mathbf{q}(t),\mathbf{q}(t');\omega]\,,
\end{equation}
for $t$, $t' \gg t_0$.

In Refs.~\cite{HZ,ABV}, the uncertainty relation of the particle of
a harmonic oscillator, coupled linearly to a bath of quantum
oscillators, is studied, and the reduced density matrix of a given
initial Gaussian wavepacket is obtained exactly. Apart from the
intrinsic dispersion due to spreading of the wavepacket, the
interaction with the environment may lead to additional dispersion
in the particle's velocity. For example, the uncertainty of either
the position or the momentum has been found to have two distinct
contributions~\cite{BA}; one of which comes from intrinsic
quantum-mechanical spreading of the wavepacket and the other is the
consequence of coupling with the quantum fluctuations of the
environment. Similar results will be shown later for the case of a
static charge.

In general, the environment-field contribution to the velocity
dispersion, denoted by $\langle\Delta v^{2}(t_{f})\rangle_{\xi}$,
will have ultraviolet divergence because we sum up all modes of
the zero-point fluctuations. Thus we have to regularize it by
introducing a cutoff frequency. The cutoff-dependent terms of the
regularized result will be used to renormalize the intrinsic
velocity dispersion caused by a finite-width wavepacket. Because
the particle's wavepacket provides a natural cutoff scale, the
contributions from environment modes with frequencies much higher
than the inverse of wavepacket size are suppressed and need not be
taken into consideration. If the cutoff scale is chosen to be the
width of the initial Gaussian wavpacket, this regularized result
will be shown to give the intrinsic velocity dispersion a
perturbative correction, which is about the order of the coupling
constant. Explicit implementation of the renormalization of the velocity dispersion and relevant discussions will be given later in the case of the static charge.

As for non-inertial motion, since the influence of the normal vacuum fluctuations on the charge's velocity dispersion has been studied in~\cite{HWL}, here we focus on the
modification of the velocity dispersion $\delta\langle\Delta
v^{2}(t_{f})\rangle_{\xi}$ due to squeezed vacuum
fluctuations of environmental modes to find any possible reduction scheme in velocity dispersion. This can be
viewed as the effects solely coming from excitations of the corresponding
modes from normal vacuum into squeezed vacuum.  The effect of the
finite wavepacket size on $\delta\langle\Delta
v^{2}(t_{f})\rangle_{\xi}$ can be argued to be negligible under a
narrow width approximation. This can be seen as follows. We assume
that the modes within a certain band are excited to the squeezed
vacuum states, and the rest of them remain in their normal vacuum
states. Now consider the band with the mean frequency $\Xi$ and the
bandwidth $\Delta$, and let the excited modes be distributed over an
narrow solid angle $d\Omega_{s}$ about a certain direction $k^i$ in
the momentum space. We denote the angular contribution over the
solid angle $d\Omega_{s}$ as
$A(d\Omega_{s})\equiv\frac{d\Omega_{s}}{(2\pi)^{3}}\left(\delta^{ii}-\frac{k^{i2}}{\omega^{2}}\right)$.
Moreover, the squeeze parameters are assumed to be mode-independent
within the band; thus their subscript $\mathbf{k}$ will be dropped
hereafter. If the initial wavepacket has an uncertainty with width
$\sigma_0$, then the wavepacket will merely modulate the
contributions from the modes with wavelength
$\lambda\lesssim\mathcal{O}(\sigma_0)$ of the environmental field.
Thus, for the the band of modes under consideration with wavelength
$c \,\Xi^{-1}\gg\sigma_0$, the finite wavepacket width has no
significant effect. Let us choose the value $\sigma_0
\sim10^{-9}\,\text{m}$ for an example. Then, as long as the central
frequency $\Xi$ is below electron's plasma frequency, the average
over the reduced density matrix in Eq.~\eqref{O} can be ignored. For
non-inertial motion, under the WKB approximation the inhomogeneous
solution to the Langevin equation is given by
\begin{equation}
	v_{i}(t)= -\frac{e}{m} \int_{t_i}^{t} \, du \, \dot K(t-u) \,
\frac{\partial}{\partial u} \xi_{i} (u)\,,
\end{equation}
where $m$ is the same renormalized mass as found in the Abraham-Dirac-Lorentz equation  and  $K(\tau)$ is the kernel function of the equation of motion (see Ref.~\cite{HWL} for detailed derivations). Hence
the modification of the velocity dispersion due to squeezed vacuum of the electromagnetic field, turned on at $t=t_0$, is obtained by averaging over the noise distribution functional, and is given by
\begin{equation}\label{E:velo_dis}
    \delta\langle\Delta v_{i}^2 (t)\rangle_{\xi}=\frac{e^2}{m^2}\int^t_{t_0} du\!\!\int^t_{t_0} du'\;\dot{K}(t-u)\dot{K}(t-u')\, \frac{\partial}{\partial u}\frac{\partial}{\partial u'}\, \delta  G_H^{ii}[\mathbf{q}(u),\mathbf{q}(u');u ,u']\,,
\end{equation}
The kernels $\delta G_{H}^{ij}$ is obtained from \eqref{Hadamrd} by subtracting out the normal vacuum contribution.

\section{velocity dispersion}\label{sec2}
\subsection{static charge}
Here we consider a static charged particle centered at $\mathbf{q}=0$ with $V(\mathbf{q})=0$. The additional wiggling in particle's trajectory due to the quantum fluctuations of the electromagnetic field can be equivalently thought of as widening or narrowing of the width of charge's wavepacket. Let us denote $\sigma_{0}$ the original,  bare width of the wavepacket before the charge comes into interaction with the quantized electromagnetic field. The initial density matrix is assumed to take the Gaussian form,
\begin{equation}
    \rho_{e}(\mathbf{q}_{1},\mathbf{q}_{2},t_i)
    =\left(\frac{1}{2\pi\sigma_{0}^{2}}\right)^{3}\exp\biggl[-\frac{(\mathbf{q}_{1}+\mathbf{q}_2)^{2}+
    (\mathbf{q}_{1}-\mathbf{q}_2)^{2}}{8\sigma_{0}^{2}}\biggr]\,.
\end{equation}
It is convenient to introduce the center-of-mass and the relative coordinates $\mathbf{\bar q}$ and $\mathbf{\bar r}$ by $\mathbf{\bar q}=(\mathbf{q}^{+} +\mathbf{q}^-)/2$ and $\mathbf{\bar r}=\mathbf{q}^+-\mathbf{q}^-$ such that $\mathbf{\bar
q}_i=(\mathbf{q}_1 +\mathbf{q}_2)/2$, $\mathbf{\bar
r}_i=\mathbf{q}_1-\mathbf{q}_2$, and $\mathbf{\bar
q}_f=(\mathbf{q}_f +\mathbf{\tilde q}_f)/2$, $\mathbf{\bar
r}_f=\mathbf{q}_f-\mathbf{\tilde q}_f$. If the particle barely moves
from the initial position, then the corresponding stochastic
coarse-grained effective action \eqref{s:xi} can be obtained as
\begin{equation}\label{e:dfdkjk}
S_{\xi} [\, \bar{\mathbf{q}}, \bar{\mathbf{r}} ;\pmb{\xi}\,] =\int dt \,
\mathbf{\bar r} \cdot \big[ - \big( m_0 +\frac{e^2}{3\pi^2} \Lambda
\big) \, \ddot{\bar{\mathbf{q}}} +\frac{e^2}{ 6 \pi^2 }
\dddot{\mathbf{\bar q}} - e\, \pmb{\xi} \big] \, ,
\end{equation}
by evaluating the kernels $G_{H}^{ij}$ and $G_{R}^{ij}$ at
$\mathbf{q}=0$. An ultraviolet frequency cutoff
$\Lambda$ is introduced. It is known that the interaction with the
electromagnetic fields leads to not only mass renormalization of the
charge, but also the self-force, given by a third-order time
derivative of the position~\cite{JA}. Essentially the noise $\xi$
accounts for the effects of the quantum fluctuations of the
environment field. We may define the renormalized mass $m= m_0
+\frac{e^2}{3\pi^2} \Lambda$ in \eqref{e:dfdkjk} and immediately see
that the correction is perturbatively small if the cutoff energy
scale $\Lambda$ is chosen to be the inverse of the wavepacket width,
because typically the width is much larger than the Compton
wavelength of the nonrelativistic particle~\cite{BRE}. Following the
same steps in \cite{Hak}, it is straightforward to obtain the exact
reduced density matrix at the final time $t_{f}$ from
Eq. \eqref{Srho},
\begin{align}\label{rhofree}
\rho_{r}(\mathbf{\bar q}_{f},\mathbf{\bar
r}_{f},t_f)&=\int\mathcal{D}\pmb{\xi}\;\mathcal{P}[\pmb{\xi}]\,\int\!d^{3}\mathbf{\bar
q}_{i} \int\ d^{3}\mathbf{\bar r}_{i}\;\left(\frac{m}{2\pi
(t_f-t_{i})}\right)^{3}
    \, \\ \nonumber
&\quad\quad\quad      \exp\biggl[\,i\,\frac{m}{t_f-t_{i}}\,(\mathbf{\bar
q}_{f}-\mathbf{\bar q}_{i}) \cdot (\mathbf{\bar r}_{f}-\mathbf{\bar
r}_{i}) +i\, e \int_{t_i}^{t_f}dt \;\mathbf{\dot{\bar r}}(t) \cdot
\mathbf{\xi}(t) \biggr]
    \times\rho_{e}(\mathbf{\bar q}_{i},\mathbf{\bar r}_{i},t_i)\,,
\end{align}
where $ \mathbf{\bar r} (t)=\mathbf{\bar r}_{i}+(\mathbf{\bar r}_{f}-\mathbf{\bar r}_{i})\times(t-t_{i}) /(t_f-t_{i})$. The velocity dispersion can be then computed from Eq.~\eqref{O}. Since the canonical momentum operator can be expressed in terms of the derivatives with respect to the position in the coordinate representation and the noise is the manifestation of the electromagnetic vector potential, the velocity operator $\pmb{v}$ is given by
\begin{equation}
 \pmb{v}(t_f)= -\frac{i}{m} \frac{\partial}{\partial \mathbf{\bar q}_f}+ \frac{e}{m}\,\pmb{\xi} (t_f) \, .
\end{equation}
With the help of \eqref{rhofree}, the velocity dispersion is given by
\begin{eqnarray}\label{E:owrpwnj}
    \langle\Delta v_{i}^2(t)\rangle_{\rm tot} &=& \langle\Delta
    v_{i}^2(t)\rangle_{\sigma} +\langle\Delta v_{i}^2(t)\rangle_{\xi}
    \nonumber \\
    &=& \frac{1}{4 m^2 \sigma_0^2}+ \frac{e^2}{m^2}\int^{t}_{t_i} du\!\int^{t}_{t_i}
    du'\; \frac{\partial}{\partial u} \frac{\partial}{\partial u'} \, G^{ii}_H (0,0;u,u')
    \, .
\end{eqnarray}
Apparently, the first term comes from the intrinsic quantum mechanical uncertainty of the particle and the second term is the consequence of the backreaction of the environmental fields, which can be written simply in terms of the $ \mathbf{E}$ field correlation function in a gauge invariant way. It is seen that an extra contribution to the velocity dispersion comes from the free field fluctuations. However, these field fluctuations might have acquired corrections from the loop effects of vacuum polarization due to virtual production and annihilation of electron-positron pairs within the context of QED. In addition, the charge $e$ would have also been renormalized by the vertex corrections of the loop effects. But presumably these loop effects should contribute a perturbative correction in the low energy regime. Thus the result of our approach by treating a nonrelativistic charged particle quantum-mechanically should give a leading-order result although a complete and systematical treatment on these effects requires the full QED calculations.

On substituting the expression of $ G^{ii}_H$ in Eq. \eqref{Hadamrd}, the induced velocity dispersion $\langle\Delta v_{i}^2(t)\rangle_{\xi} $ can be further split into the part due to the zero-point fluctuations of the fields $\langle\Delta v_{i}^2(t)\rangle_{\xi,\,0}$ and the part from the fluctuations of excited squeezed modes $\delta\langle\Delta v^2_i(t)\rangle_{\xi}$,
\begin{equation}
\langle \Delta v_{i}^2(t)\rangle_{\xi}=\langle\Delta
v_{i}^2(t)\rangle_{\xi,\,0}+\delta\langle\Delta
v^2_i(t)\rangle_{\xi} \, .
\end{equation}
The expression $\langle\Delta v_{i}^2(t)\rangle_{\xi,\,0}$ is formally divergent and is saturated to a cutoff-dependent constant as $t_i \to -\infty$,
\begin{equation}
 \langle\Delta v_{i}^2(t)\rangle_{\xi,\,0} = \frac{e^2}{m^2} \frac{1}{3 \pi^3} \int^{\Lambda} d\omega\; \omega \Bigl\{ 1-\cos[\omega (t-t_i)] \Bigr\} =\frac{e^2\Lambda^2}{6\pi^2  m^2} \Bigl\{ 1+\mathcal{O}\Bigl(\Lambda^{-1}(t-t_i)^{-1}\Bigr) \bigr\}\,.
\end{equation}
Because we can never turn off the
electromagnetic zero-point fluctuations, a free charged particle is always affected by them. Thus this cutoff-dependent result can be absorbed into the intrinsic
quantum mechanical uncertainty such that the intrinsic velocity
uncertainty acquires a correction. If we choose the cutoff to be of
the order $\sigma_0^{-1}$, the inverse of the initial width of the
wavepacket, then the renormalized width is given by $
\sigma=\sigma_0 -\frac{e^2}{3 \pi^2}\,\Lambda^2 \sigma_0^3$, with a
perturbatively small correction of the order of the fine structure
constant. Similar cutoff-dependence effects to the corrections of
the width are also found in \cite{BA}.

\begin{figure}
\centering
    \scalebox{0.700}{\includegraphics{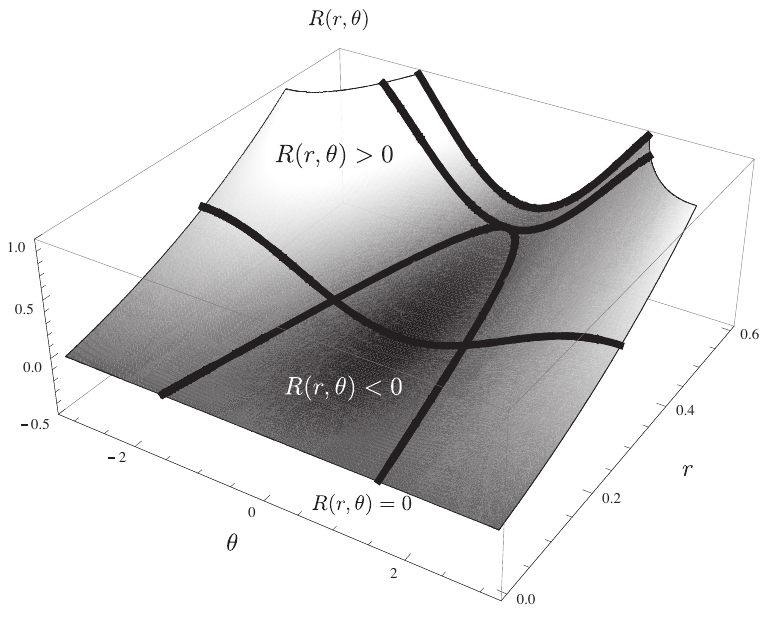}}
    \caption{The surface plot of $R(r,\theta)$.}\label{Fi:dksl}
\end{figure}

Hence from hereafter we will
identify $m$ and $\sigma$ as the observed physical mass and width of
the particle state to be determined experimentally. In this article, we consider
the term $\delta\langle\Delta v^2_i(t)\rangle_{\xi}$, the
modification of the renormalized velocity
dispersion due to the excited squeezed vacuum modes. In a similar idea to the Casimir effect, it is not the effect
due to Minkowski vacuum to be observed, but the effect due to
variations of Minkowski vacuum, given by the change of the states or
configurations, to be measured. To be more precise, we consider
that only a finite band of the electromagnetic modes are excited to
the squeezed vacuum states while the rest of the modes remain in
their normal vacuum states. Let $\Xi$ and $\Delta$ be the mean
frequency and width of the band, respectively, and suppose the wave
vectors are distributed over small solid angle $d\Omega_{s}$ about a
certain direction. Thus $\delta\langle\Delta v^2_i(t)\rangle_{\xi}$
is given by
\begin{equation*}
    \delta\langle\Delta v^2_i(t)\rangle_{\xi}=\frac{e^2}{m^2}A(d\Omega_{s})\int_{\Xi-\Delta/2}^{\Xi+\Delta/2}\!\!d\omega\;4\omega\Bigl[\eta^{2}+\mu\eta\,\cos(\omega t-\theta)\Bigr]\, \sin^2 \frac{\omega t}{2} \,.
\end{equation*}
where the time $t_0$ of turning on the squeezed modes are set to be $t_0=0$, and the factor $A(d\Omega_{s})$ describes the angular contribution of excited modes. Let us assume that the squeezed parameters $\mu$, $\eta$ and $\theta$ barely change with frequency within the band. Carrying out the $\omega$-integral gives
\begin{align*}
    &\quad\int_{\Xi-\Delta/2}^{\Xi+\Delta/2}d\omega\;\omega\,\eta\Bigl[\eta+\mu\,\cos(\omega t-\theta)\Bigr]\sin^2\frac{\omega t}{2}\\
    &=\frac{\Delta\Xi}{4}\,\eta\Bigl[2\eta-\mu\,\cos\theta\Bigr]+\frac{\eta}{8t^2}\sin\frac{\Delta }{2}t\left[8\eta\,\sin\Xi t-8\mu\,\sin(\Xi t-\theta)+2\mu\,\cos\frac{\Delta}{2}t\,\sin(2\Xi t-\theta)\right]\\
    &\qquad-\frac{\eta}{8t}\left\{\left[\Xi+\frac{\Delta}{2}\right]\left[\mu\,\sin(2\Xi t+\Delta t-\theta)-4\mu\,\sin(\Xi t+\frac{\Delta}{2} t-\theta)+4\eta\,\sin(\Xi t+\frac{\Delta}{2} t)\right]\right.\\
    &\qquad\qquad\left.-\left[\Xi-\frac{\Delta}{2}\right]\left[\mu\,\sin(2\Xi t-\Delta t-\theta)-4\mu\,\sin(\Xi t-\frac{\Delta}{2} t-\theta)+4\eta\,\sin(\Xi t-\frac{\Delta}{2} t)\right]\right\}\,.
\end{align*}
Except for the expression in the first pair of square brackets, the rest reveal a power-law decay in time. We find, at asymptotic times the change of the renormalized velocity dispersion is dominated by the constant
\begin{equation}\label{E:dfkjs}
  \delta\langle\Delta v^2_i(t)\rangle_{\xi}=\frac{e^2}{m^2} \, A(d\Omega_{s}) \left(2 \Xi \Delta\right)\left[\eta^2-\frac{1}{2}\,\mu\eta\,\cos\theta\right]\,.
\end{equation}
The nonstationary component of $\delta\langle\Delta v^2_i(t)\rangle_{\xi}$, derived from the corresponding component of the noise-noise correlation function, exhibits an interesting feature that it may change the sign with an appropriate choice of squeeze parameters $r$ and $\theta$. The expression in the squared brackets of Eq.~\eqref{E:dfkjs} is always greater than $\eta^{2}-\mu\eta/2$, which is bounded below,
\begin{equation}\label{E:lowerbouded}
    \eta^{2}-\frac{1}{2}\mu\eta\geq-\frac{2-\sqrt{3}}{4}>-\frac{1}{2}\,.
\end{equation}
Thus, the change of the renormalized velocity
fluctuations can be negative. In particular, it implies that this
observed velocity dispersion in squeezed vacuum can be reduced to a level lower than its counterpart in normal vacuum of the electromagnetic fields with the same
frequency band. However, the reduction has a lower bound. Further discussion would be more transparent if we single out the
corresponding velocity dispersion $\langle\Delta
v^2_i(t)\rangle_{\xi,\,0}$ due to normal vacuum. It is given by replacing $\eta^2$
with $1/2$ and $\mu$ with $0$ respectively in Eq.~\eqref{E:dfkjs},
\begin{equation}\label{E:sddfkjs}
  \langle\Delta v^2_i(t)\rangle_{\xi,\,0}= \frac{e^2}{m^2} \, A(d\Omega_{s}) \left(2 \, \Xi \Delta\right)\times\frac{1}{2}\,.
\end{equation}
We may then compare the value of $\delta\langle\Delta
v^2_i(t)\rangle_{\xi}$ with $\langle\Delta
v^2_i(t)\rangle_{\xi,\,0}$.
Let the ratio of $\delta\langle\Delta v^2_i(t)\rangle_{\xi}$ to $\langle\Delta v^2_i(t)\rangle_{\xi,\,0}$ define a function $R(r,\theta)$,
\begin{equation}
    R(r,\theta)=2 \eta^2-\mu\eta\cos\theta\,,
\end{equation}
whose dependence on the squeeze parameters is shown in Fig.~\ref{Fi:dksl}. The function $R(r,\theta)$ is negative in the region shaded in darker gray and encircled by the curve $R(r,\theta)=0$. Three contours of $R(r,\theta)$ with different fixed values of $r$ are highlighted by thick solid curves, among which the foremost one passes through the minimum of $R(r,\theta)$. It is clearly seen that with larger values of $r$, the interval of $\theta$ where $R(r,\theta)$ is negative becomes increasingly narrower.
From Eq.~\eqref{E:lowerbouded}, we see that the
minimum value of $\eta^{2}-\mu\eta/2$ is still greater than $-1/2$. This can be understood by the fact that the total
velocity dispersion, the sum of Eqs. \eqref{E:dfkjs} and
\eqref{E:sddfkjs}, is always positive, as is required by the
definition of the velocity dispersion. More general inequality  can
be found in Appendix~\ref{appenb}. Hence, the velocity dispersion
can be maximally reduced by a factor $\sqrt{3}/2\sim0.866$ from its
counterpart solely due to the normal vacuum fluctuations. The
corresponding photon number is $\bar{n}=\eta^2=(2\sqrt{3}-3)/6\ll1$,
namely, there is less than one photon in each mode $\mathbf{k}$.

So far we have shown a possible scheme of reducing the velocity
dispersion of the charge by the nonstationary noises, which are
manifested from the electromagnetic squeezed vacuum fluctuations.
This is an unusual feature that deserves further discussion. Let us
consider that the charge comes into contact with squeezed vacuum
fluctuations at some nonvanishing value of $t_0$. Eq.~\eqref{E:dfkjs} then becomes
\begin{equation}\label{E:dfkjs-t0}
  \delta\langle\Delta v^2_i(t)\rangle_{\xi}=\frac{e^2}{m^2} \, A(d\Omega_{s})
  \left(2 \, \Xi \Delta\right)\left[\eta^2-\frac{1}{2}\,\mu\eta\,\cos(2\Xi\,t_0-\theta)\right]\,,
\end{equation}
and reduction may occur when $\mu\eta\,\cos(2\Xi\,t_0-\theta)>0$. In
analogy to the previous arguments, maximal reduction can be achieved by choosing suitable values of the
squeezing parameter. Nevertheless, as is expected, any uncertainty
associated with the initial time $t_0$ may give an average-out
effect to undermine this noise-reduction mechanism. The averaged
result over fairly large uncertainty will instead lead to
\begin{equation}\label{E:dfkjskq}
  \delta\langle\Delta v^2_i(t)\rangle_{\xi}=\frac{e^2}{m^2} \, A(d\Omega_{s})
  \left(2 \, \Xi \Delta\right)\;\eta^2>0\,,
\end{equation}
in which the change in the velocity dispersion is solely determined
by the amplitude of the squeezing parameter $\eta$, that is, by the
number of the photons, $\bar{n}=\eta^2$. Thus, roughly speaking, as
long as the uncertainty of the time $t_0$ is limited to a
value smaller than the inverse of central frequency of squeezed
light $\Xi$, noise reduction will be possibly observed.

The same line of thought on an electromagnetic squeezed vacuum has been applied to restore coherence in electron interferometry~\cite{HF}. There, loss of coherence in the electron interference experiment can be partially restored if the background electromagnetic field is prepared in its squeezed vacuum state. This requires that electrons should be emitted at a selected interval within the cycle of the excited modes of the electromagnetic field to form interference fringe. The effect of recovering coherence between electron partial waves is constrained by an inequality, and once all electrons are allowed to interfere, decoherence reappears.

Next, let us reformulate the uncertainty of the initial time, in terms of a switching function, as a consequence of the finite-time switching-on/off process of interaction between the charged particle and electromagnetic squeezed modes.

\subsection{finite-time switching process}
The finite-time switching effect can be incorporated by introducing an appropriate switching function. Here a rather general function $f$ with two time scales $\tau$ and $t$, which describe the switching processes and the measurement duration respectively, are chosen,
\begin{equation}\label{E:wpoms}
    f(u)=\begin{cases}
            e^{u/\tau}\,;& -\infty < u<0\,,\\
            1\,;&0<u<t\,,\\
            e^{-(u-t)/\tau}\,;&t<u < \infty\,.
        \end{cases}
\end{equation}
The sudden switching process in the previous section corresponds to the limit $\tau\to0$.  Then the change in the velocity dispersion for a static charge is described by 
\begin{equation}
    \delta \langle\Delta v_{i}^2(t)\rangle_{\xi}=\frac{e^2}{m^2}\int^{\infty}_{-\infty} du \int^{\infty}_{-\infty}
    du'\; f(u)f(u')  \,\frac{\partial}{\partial u}\frac{\partial}{\partial u'}\, \delta G^{ii}_H (0,0;u,u')\,.
\end{equation}
We will set $t_{0}=0$ in all subsequent sections. With an aid of Appendix~\ref{appenb}, an inequality of the velocity dispersion in terms of this switching function can be found to be
\begin{align}
    \delta\langle\Delta
    v^2_i(t)\rangle_{\xi}&\geq-\frac{e^2}{m^2}\,A(d\Omega_s)\int_{\Xi-\frac{\Delta}{2}}^{\Xi+\frac{\Delta}{2}}\!\!d\omega\;\frac{\omega}{2}\int_{-\infty}^{\infty}\!du\!\int_{-\infty}^{\infty}\!du'\;f(u)f(u')\,e^{-i\omega(u-u')}\notag\\
    &=-\frac{e^2}{m^2}\,A(d\Omega_s)\int_{\Xi-\frac{\Delta}{2}}^{\Xi+\frac{\Delta}{2}}\!\!d\omega\;\frac{\omega}{2}\left|\int_{-\infty}^{\infty}\!du\;f(u)\,e^{-i\omega u}\right|^{2}\,.\label{E:owiek}
\end{align}
It implies a lower bound on the negative value of the modification of the velocity dispersion of a static charge. The Fourier transform of the switching function in the frequency domain is
\begin{equation}
    \int_{-\infty}^{\infty}\!du\;f(u)\,e^{-i\omega\,u}=\frac{1}{\omega}\left(\frac{1}{i+\omega\tau}+\frac{e^{-i\omega
    t}}{-i+\omega\tau}\right)\,.
\end{equation}
Then the inequality takes the form
\begin{equation} \label{lowerb}
     \delta\langle\Delta v^2_i(t)\rangle_{\xi} \geq -\frac{e^2}{m^2}\,A(d\Omega_s)\Bigl[g(\tau,t;\Xi+\frac{\Delta}{2})-g(\tau,t;\Xi-\frac{\Delta}{2})\Bigr]\,,
\end{equation}
where
\begin{equation*}
    g(\tau,t;\omega)=\int^{\omega}\!\!d\omega'\;\dfrac{2\omega'\left[\omega'\tau\cos(\frac{\omega't}{2})+\sin(\frac{\omega't}{2})\right]^2}{\left(1+\omega'^2\tau^2\right)^2}\,.
\end{equation*}
In particular, if the measuring duration $t$ is sufficiently long, we may let $t\to\infty$, and  see that the function $g(\tau,t;\omega)$ is saturated to the value,
\begin{equation*}
   g(\tau,t=\infty;\omega)=
   \frac{1}{2\tau^2}\ln\left[1+\omega^2\tau^2\right] \, .
\end{equation*}
Thus, the lower bound in Eq.~\eqref{lowerb} crucially depends on the time scale $\tau$. Two asymptotic switching processes will be considered. The sudden switching corresponds to $\tau\ll1/\omega$ while the adiabatic switching means $\tau\gg1/\omega$, if the typical frequency of the modes in the band be $\omega$. Then the  quantity $g(\tau,t=\infty;\omega)$ in these limits takes the respective values
\begin{equation}
    g(\tau,t=\infty;\omega)=\begin{cases}
            \frac{\omega^2}{2}+\mathcal{O}(\tau^2)\,;&\tau\ll \omega^{-1}\,,\\
            \frac{\ln(\omega\tau)}{\tau^2}+\mathcal{O}(\tau^{-3})\,;&\tau\gg \omega^{-1}\,.
        \end{cases}
\end{equation}
So for an adiabatic switching, \eqref{lowerb} approaches zero as
\begin{equation}
    \delta\langle\Delta
v^2_i(t)\rangle_{\xi}\geq-\frac{e^2}{m^2}\,A(d\Omega_s)\,\frac{1}{\tau^2}\,\ln
\left[\frac{\Xi-\frac{\Delta}{2}}{\Xi+\frac{\Delta}{2}}\right]^2 \,.
\end{equation}
It means that in such a limit we cannot effectively suppress the
velocity dispersion of the charge in a slow switching process. On the other hand, for a sudden
switching, the inequality gets back to
\begin{equation}
   \delta\langle\Delta
 v^2_i(\infty)\rangle_{\xi}\geq-\frac{e^2}{m^2}\,A(d\Omega_s)\,\Xi\Delta\,,
\end{equation}
consistent with the result in \eqref{E:lowerbouded} and \eqref{E:sddfkjs}. Thus, in order to have the optimal suppression on the velocity dispersion of the particle, the charged particle (system) should come into interaction with the properly chosen quantum states of the environmental field within a time scale much shorter than any other scale in the problem. A slow switching process will hinder the effect of noise reduction on the velocity dispersion of the particle.

\subsection{non-inertial motion of a charge}
Now consider the center of the wavepacket undergoes a simple harmonic motion with angular frequency $\omega_0$ in the direction $i$ with $V(\mathbf{q}) =\frac{1}{2}\,m_0 \omega_0^{2}q_i^2$. The modification of the velocity dispersion of the charge due to the electromagnetic squeezed vacuum fluctuations is described by
\begin{equation}\label{E:vel_dis}
    \delta\langle\Delta v_{i}^2 (t)\rangle_{\xi}=\frac{e^2}{m^2}\int^t_0du\!\!\int^t_0du'\;\dot{K}(t-u)\dot{K}(t-u')\, \delta G_H^{ii}[\mathbf{q}(u),\mathbf{q}(u');u ,u']\,,
\end{equation}
Thus, this kernel function $K(\tau)$ serves to dynamically modulate the quantum fluctuations of the environment fields, and in turn modify its influences on the evolution of the velocity dispersion along the trajectory of the center of the wavepacket. The time derivative of the kernel function $\dot{K}$~\cite{HWL} is
\begin{equation}
    \dot{K}(\tau)=Z\,e^{-\Gamma\tau}\cos(\Omega\,\tau+\delta)\,,
\end{equation}
with the resonance frequency $\Omega$ and the decay constant $\Gamma$ given by~\cite{HWL}
\begin{align}\label{E:Omegamma}
    \Omega&\sim\omega_0+\frac{\operatorname{Re}\Sigma(\omega_0)}{2\omega_0}\,,&Z&\sim\left[1-\frac{\partial\operatorname{Re}\Sigma(\Omega)}{\partial\Omega^2}\right]^{-1}\,, \nonumber \\
    \Gamma&\sim Z\,\frac{\operatorname{Im}\Sigma(\Omega)}{2\Omega}\,,&\delta&\sim Z\,\frac{\partial\operatorname{Im}\Sigma(\Omega)}{\partial\Omega^2}\,.
\end{align}
Thus, the dissipation backreaction, which is given by the electromagnetic self-force due to this non-inertial motion, renders the trajectory into underdamped motion. The kernels $\operatorname{Re}\Sigma(\omega)$ and $\operatorname{Im}\Sigma(\omega)$ are defined by
\begin{equation*}
    \widetilde{\Sigma}(s=i\,\omega+0^{\pm})=\operatorname{Re}\Sigma(\omega)\pm i\operatorname{Im}\Sigma(\omega)\,,
\end{equation*}
and
\begin{equation*}
    \widetilde{\Sigma}(s)=\frac{e^2}{m}\,s^2\left\{\widetilde{G}_{R}^{ii}[\mathbf{q}=\mathbf{q}'=0;s]-\frac{\Lambda}{3 \pi^2} \right\}\,,
\end{equation*}
where the function $\widetilde{G}_{R}^{ii}[\mathbf{q}=\mathbf{q}'=0;s]$ is
the Laplace transformation of the retarded Green's function in the
dipole approximation, $e^{i\mathbf{k}\cdot\mathbf{q}}\approx1$.
Under this situation, the semiclassical approximation gives an exact
result of the velocity dispersion from the influence of the
environmental fields~\cite{ABV}. The ultraviolet divergence is
absorbed by appropriate mass renormalization of the particle with
the physical mass $m$ given by $m= m_0 +\frac{e^2}{3\pi^2} \Lambda
$, in which $\Lambda$ is an ultraviolet cutoff frequency. Hence, up
to order $e^2$, we obtain~\cite{HWL}
\begin{equation}\label{E:skfjsk}
    \operatorname{Re}\Sigma(\omega)=0\,,\qquad\qquad\operatorname{Im}\Sigma(\omega)=\frac{e^2}{4\pi m}\,\operatorname{sgn}(\omega)\,\frac{ 2\omega^3}{3}\,,
\end{equation}
and in the weak coupling limit, we have $\Gamma\ll\Omega$. Then, Eq.~\eqref{E:vel_dis} can be further simplified to,
\begin{align}
    \delta\langle\Delta v^2_i(t)\rangle_{\xi,\,\text{st}}&=\frac{e^2}{m^2}A(d\Omega_{s})\,\eta^{2}\int_{\Xi-\Delta/2}^{\Xi+\Delta/2}\!d\omega\,\frac{\omega^{3}}{2}\!\int^t_0\!\int^t_0\!d\tau d\tau'\,\dot{K}(\tau)\dot{K}(\tau')\,e^{-i\omega(\tau-\tau')}+\text{c.c.}\,,\label{E:sdfsk}\\
    \delta\langle\Delta v^2_i(t)\rangle_{\xi,\,\text{ns}}&=\frac{e^2}{m^2}A(d\Omega_{s})\,\mu\eta\int_{\Xi-\Delta/2}^{\Xi+\Delta/2}\!d\omega\,\frac{\omega^{3}}{2}\!\int^t_0\!\int^t_0\!d\tau
    d\tau'\,\dot{K}(\tau)\dot{K}(\tau')\,e^{i\theta-i\omega(2t-\tau-\tau')}+\text{c.c.}\,.\label{E:dkjfhs}
\end{align}
They respectively represent the the stationary and the nonstationary components of the change in the velocity fluctuations as a consequence that a band of the field modes are excited to squeezed vacuum states from the corresponding normal vacuum of the electromagnetic fields.

Here we will consider the narrow band case, $\Delta\ll\Xi$. The extension of our study to the broad band case is straightforward if more details about the mode dependence of the squeeze parameters are known. In principle, since each mode influences the charge's dynamics independently, we may split the broad band into many narrow bands within which the squeeze parameters are frequency-independent but may take different values. Thus the result of the broadband case with frequency-dependent squeeze parameters amounts to adding up effects from all the narrow bands. Hence the narrow band example is simple enough to have analytical expressions, but sufficiently sophisticated to capture all essences.

We assume that the resonance frequency $\Omega$ of the motion lies within the band of the squeezed vacuum modes. For simplicity, let the mean frequency of the band $\Xi$ coincide with the resonance frequency $\Omega$, namely $\Xi=\Omega$, and consider the situation that the bandwidth $\Delta$ is much smaller than the mean frequency $\Xi$. Recall that $\Gamma=(e^2/12\pi m)\,\Omega^2 $, as seen from Eqs.~\eqref{E:Omegamma} and~\eqref{E:skfjsk}. Then $\Gamma/ \Omega \sim r_c\times\Omega/c$, where the classical radius of the charge $r_{c}$ is defined by $r_c=e^2/4\pi m$. If we choose the charge as an electron, we have $r_c\sim2.82\times10^{-15}\,\mathrm{m}$ and then
\begin{equation}
    \frac{\Gamma}{\Delta}\approx10^{-17}\;\frac{\Delta}{\Omega}\;\frac{\Omega}{10^6\mathrm{s}^{-1}}\,,
\end{equation}
which is an extremely small value for the typical choice of $\Omega$ and $\Delta$. Hence, $\Gamma\ll\Delta$ is a plausible assumption. We then assume that $\Gamma\ll\Delta\ll\Xi=\Omega$ holds for the following discussion.

Integration over $\tau$ and $\tau'$ variables in Eqs.~\eqref{E:sdfsk} and \eqref{E:dkjfhs} is carried out in Appendix~\ref{appendc}. When the particle sets into oscillatory motion, we notice that the integrands of the $\omega$-integral in Eqs.~\eqref{E:stw} and \eqref{E:nsw} have a Breit-Wigner feature of the narrow resonance due to weak coupling. The resonance peaks at frequency $\Omega$ and has width $\max\{2\pi/t,\Gamma\}$. It indicates that the resonance width decreases with time as $t^{-1}$ and then approaches to the value $\Gamma$ as the motion evolves into the relaxation regime. At early time $\Omega^{-1}\ll t\ll\Delta^{-1}$, since the resonance width is of the order $t^{-1}$, it is greater than the bandwidth of the squeezed vacuum modes $\Delta$, that is, $\Delta\,t\ll1$. The integrands in Eqs.~\eqref{E:stw} and \eqref{E:nsw} thus slowly vary inside the bandwidth, and can be pull out of the $\omega$-integral by substituting $\omega$ with $\Xi$. For narrow resonance under consideration, the most dominant contributions of the stationary component come from $L_1(\Omega)$ and $L_2(\Omega)$ in Appendix~\ref{appendc} because the sum of them contains terms of the form $[(\omega-\Omega)^{2}+\Gamma^{2}]^{-1}$ around the resonance frequency. Similarly, the nonstationary component is dominated by the term $J_{1}(\Omega)$ since it varies like $(\omega-\Omega)^{-2}$ in the neighborhood of the resonance peak. By combining the stationary and nonstationary components of $\delta\langle\Delta v_i^2(t)\rangle_{\xi}$, we find that, at early time $\Omega^{-1}\ll t\ll\Delta^{-1}$, the modification of the velocity dispersion is given by
\begin{equation}\label{E:dwie}
    \delta\langle\Delta v_i^2(t)\rangle_{\xi} \simeq \frac{e^2}{m^2} \, A(d\Omega_{s})  \biggl(\frac{1}{4} \, \Delta\Omega^3 \biggr)\biggl[\eta^{2}+\mu\eta\,\cos(2\Omega t-\theta+2\delta)\biggr]\,t^2\,,
\end{equation}
which grows quadratically in time, and depends on the bandwidth of the squeezed vacuum modes $\Delta$.

As the time progresses to the regime $\Delta^{-1}\ll t\ll\Gamma^{-1}$, the resonance width gradually decreases to a value about the same order of magnitude as $\Gamma$, which is smaller than the bandwidth $\Delta$. It implies there may exist a transition or crossover of the time dependence of $\delta\langle\Delta v_i^2(t)\rangle_{\xi}$ at about $t\simeq\Delta^{-1}$. In addition, we expect when $t\gg\Delta^{-1}$, the squeezed modes may start to evolve out of phase with each other. This may lead to cancelation between modes in the contributions of the nonstationary component, and may slow down the growth of the corresponding component of velocity dispersion. Analytically we find that the stationary component of $\delta\langle\Delta v_i^2(t)\rangle_{\xi}$ in this time regime, $\Delta^{-1}\ll t\ll\Gamma^{-1}$, increases linearly in time,
\begin{equation}\label{E:gsdgde}
    \delta\langle\Delta
    v_i^2(t)\rangle_{\xi,\,\text{st}}\simeq \frac{e^2}{m^2} \,A(d\Omega_{s})\, \eta^2 \, \biggl(\frac{\pi}{2}\,\Omega^3\,t\biggr)\,.
\end{equation}
Heuristically this evolution behavior can also be obtained from the stationary component of Eq.~\eqref{E:dwie} by replacing the bandwidth $\Delta$ with the resonance width, which is of order $t^{-1}$. This can be understood by the fact that when the bandwidth is wider than the resonance width, the result of  velocity dispersion should not explicitly depend on bandwidth~\cite{HWL}. In contrary, the nonstationary component flattens out, and behaves like
\begin{equation}\label{E:sgrfdgde}
    \delta\langle\Delta
    v_i^2(t)\rangle_{\xi,\,\text{ns}}\simeq \frac{e^2}{m^2} \,A(d\Omega_{s}) \, \mu\eta \left(\frac{\Omega^3}{\Delta}\right)\,\cos(2\Omega\,t-\theta+2\alpha)\,.
\end{equation}
It is seen that the growth rate of the envelope of the nonstationary component starts falling behind that of the stationary component since coherence between the squeezed vacuum modes is gradually lost. This cancelation effect will be more significant as the evolution moves into the relaxation regime.
\begin{figure}
\centering
    \scalebox{1}{\includegraphics{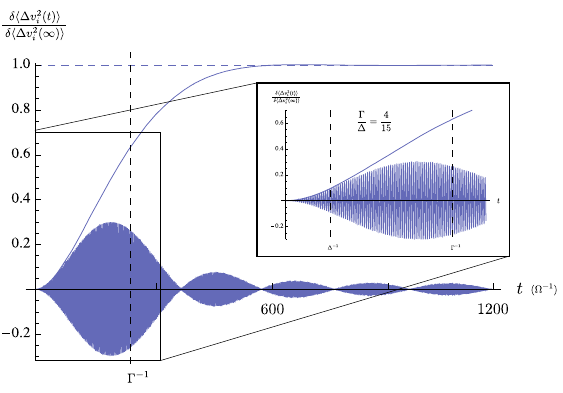}}
    \caption{The full-time evolution of the change of the velocity dispersion of a wavepacket with its center undergoing simple harmonic motion in the case $\Gamma\ll\Delta\ll\Omega=\Xi$  with $\Gamma/\Omega=0.004$ and $\Delta/\Omega=0.015$ is drawn. The nonstationary component of $\delta\langle\Delta  v_i^2(t)\rangle_{\xi}$ oscillates rapidly and then vanishes eventually. The stationary component grows at early time and reaches saturation at asymptotical times.}\label{Fi:sdsl}
\end{figure}

Finally, at much later time, $t\gg\Gamma^{-1}$, the nonstationary component falls off with time as $t^{-1}$, and vanishes eventually
\begin{equation}\label{widelate}
    \delta\langle\Delta v_i^2(t)\rangle_{\xi,\,\text{ns}}\simeq-\frac{e^2}{m^2} \, A(d\Omega_{s})\,\mu\eta \, \left(\frac{\Omega^3}{\Delta^{2}t}\right)\sin(\Delta\,t)\cos(2\Omega\,t-\theta+2\alpha)\,,
\end{equation}
but the stationary component of $\delta\langle\Delta v_i^2(t)\rangle$ saturates to a time-independent constant. The saturated value is given by
\begin{equation}\label{E:skhfso}
    \delta\langle\Delta v_i^2(t)\rangle_{\xi,\,\text{st}}\simeq \frac{e^2}{m^2}\, A(d\Omega_{s})\, \eta^{2}\, \frac{\pi}{4}
    \left(\frac{\Omega^3}{ \Gamma}\right)   \simeq \eta^2\,\frac{3\pi^2}{2}\,A(d\Omega_{s}) \,  \hbar \Omega \,.
\end{equation}
To arrive at Eq.~\eqref{E:skhfso}, we have again made an substitution of $\Gamma=(e^2/12\pi m)\,\Omega^2$. Since the nonstationary component vanishes at asymptotical times, the velocity dispersion can not be possibly reduced by manipulating the nonstationary component via squeeze parameters, and its values is solely determined by the stationary component.

It has be seen that the fluctuation-dissipation relation plays a role in order to dynamically stabilize the value of the velocity dispersion of the particle in a fluctuating environment. In particular, when the charged particle undergoes non-inertial motion, it experiences dissipation backreaction by the electromagnetic self-force, but also its velocity dispersion acquires an additional contribution from the accompanying field fluctuations. This is again a consequence of  the fluctuation-dissipation relation. Thus, it is no surprise that velocity dispersion is saturated at late times.

The full-time evolution of $\delta\langle\Delta v_i^2(t)\rangle_{\xi}$ of a particle with the typical choice of the value of the parameters is shown in Fig.~\ref{Fi:sdsl}. The evolution of $\delta\langle\Delta v_i^2(t)\rangle_{\xi}$ in the time regime $t\ll\Gamma^{-1}$ has been zoomed up. The vertical axis is normalized by the late-time values of the change of velocity fluctuations $\delta\langle\Delta v_i^2(\infty)\rangle_{\xi}$. It is seen that the nonstationary component reveals a much faster oscillatory behavior as compared with the stationary component. The time scale after which the nonstationary component starts to die out is determined by $t\simeq\Delta^{-1}$ when the modes inside the band evolve out of coherence. Thus, the nonstationary noise gives a transient effect on the dynamics of  the velocity dispersion, and its effect depends on the motion of the charged particle. We may further observe that in comparison with $\delta\langle\Delta v_i^2(t)\rangle_{\xi,\,\text{st}}$, the values of $\delta\langle\Delta v_i^2(t)\rangle_{\xi,\,\text{ns}}$ can be significant only in the much earlier stage of motion where the evolution time is shorter than the oscillation period $t\ll\Omega^{-1}$.

\section{Summary and conclusion}\label{sec3}
The time evolution of the change of the velocity dispersion of a charged
particle, coupled to the squeezed vacuum modes of the
electromagnetic field, is studied. We find that for appropriate choices of the squeeze
parameters, the presence of the nonstationary noise may reduce the
renormalized velocity dispersion of the static charge. The maximal
reduction in terms of the effective temperature is
\begin{equation}\label{E:dfhkdhfd}
    \delta T_{\text{eff}}\sim \frac{\sqrt{3}}{4}\, \left(\frac{e^2\Xi}{m \, c}\right)\left(\frac{\Delta}{\Xi}\right)\,A(d\Omega_{s})\,\frac{\hbar \, \Xi}{k_{B}}\sim  10^{-10} \,A(d\Omega_{s})\,\left(\frac{\Delta}{\Xi}\right)\,  \left(\frac{\Xi}{10^{12}\,\mathrm{s}^{-1}}\right)^{2}\mathrm{K}\,,
\end{equation}
where $\hbar$ and $c$ have been put back and $k_B$ is the Boltzmann constant. Thus, this temperature reduction is small when the mean frequency of the squeezed vacuum modes $\Xi$ lies within the radio frequency, and the bandwidth $\Delta$ is of the same order as its mean value~\cite{MN}. A more precise quantitative evaluation requires quantum electrodynamics and it deserves further study.

An quantum inequality to constrain negativeness of the change in the velocity dispersion of the particle is derived by introducing a switching function. This switching function can describe the time scale on which the system comes into interaction with the environment as well as the measurement time scale. For large measuring time, it is shown that the lower bound of the modification of the velocity dispersion can be optimally achieved by a sudden switching process. On the other hand, reduction of quantum noise from the environment may not be as effective in a slow switching process.

When the center of the wavepacket undergoes oscillatory motion, we consider the frequency of the charged oscillator lies within the band of the squeezed vacuum modes $\Delta$. We find that the change in the velocity fluctuations in general grows with time at early moments, $t\ll\Gamma^{-1}$. In this time regime, both the stationary component and the envelope of the nonstationary component initially increase in a similar fashion, but their evolution behaviors go through a transition at $t\simeq\Delta^{-1}$ when the excited squeezed vacuum modes gradually evolve out of phase with each other. It is shown that after the transition time $\Delta^{-1}$, the stationary component does not grow as fast as at earlier time, but the nonstationary component falls off much more quickly so its envelope flattens out in this time regime. Thus the nonstationary component has become less significant than the stationary component. At late time $t\gg\Gamma^{-1}$, the nonstationary component vanishes like $t^{-1}$ but the stationary component saturates. Therefore, the change in velocity fluctuations will reach a time-independent constant, entirely determined by its stationary component. It also indicates that the squeezed vacuum fluctuations are not as effective in reducing velocity dispersion of the particle at late time. The effectiveness thus depends on the state of the motion of the wavepacket. The corresponding modification in the effective temperature is given by
\begin{equation}\label{E:dfhkdhf}
    \delta T_{\text{eff}}\sim\eta^2\,\frac{3\pi^2}{2}\,A(d\Omega_{s})\,\frac{\hbar\Omega}{k_{B}}\sim  10^{-3} \, \bar{n}\,A(d\Omega_{s})
    \left(\frac{\Omega}{10^{6}\,\mathrm{s}^{-1}}\right)\mathrm{K}\,.
\end{equation}
Compared with the result in Ref.~\cite{HWL} where the normal vacuum states of the electromagnetic fields are considered, the effective temperature we obtain here is found to depend on not only the oscillation frequency of motion, but also the mean number of photons $\bar{n}=\eta^2$ in each squeezed vacuum mode. The non-inertial motion of the particle results in dissipation backreaction in the form of the electromagnetic self-force, which in turn induces fluctuations back to the charge and then increase its velocity dispersion. This is a consequence of the underlying fluctuation-dissipation relation.

In electrodynamics, the dynamics of a charged particle is governed by coupling between the transverse component of the vector potential and the charged current density. This interaction, depending on the first-order time derivative of the particle's position, gives rise to the electromagnetic self-force, which is a third-order time derivative of the position, and it results in so-called supraohmic dynamics~\cite{CAL,HPZ}. This self-force can be argued to be insignificant in the course of the evolution of a charge in inertial motion~\cite{HPZ,WHL}. It is in striking contrast to the Brownian motion in an ohmic environment, characterized by the dissipation backreaction which is the first-order time derivative of the particle's position. This dissipative dynamics can be formulated in terms of coordinate coupling of the particle with the environment~\cite{CAL,HPZ}. Furthermore stronger dissipation should be expected to occur in the subohmic case~\cite{HPZ}. According to the fluctuation-dissipation relation, the effect of the fluctuations backreaction from the subohmic environment on the dynamics of a particle should be rather different. It is then of interest to extend the scope of the current study to compare the subvacuum effects from a fluctuating subohmic, ohmic, or supraohmic environments on the particle when the particle undergoes quantum Brownian motion, an important paradigm of quantum open systems. This work is in progress.

\begin{acknowledgments}
We would like to thank Larry H. Ford for carefully reading this manuscript. This work was supported in part by the National Science Council, R. O. C. under grant NSC97-2112-M-259-011-MY3, and the National Center for Theoretical Sciences, Taiwan.
\end{acknowledgments}

\appendix
\section{Outlines of the derivation of $S_{\mathbf{\xi}}[\mathbf{q}^{+},\mathbf{q}^{-}
    ;\xi]$}\label{appena}
From Eq.~\eqref{E:ddfff} and the definition of the propagating function \eqref{E:dfdfdfdf}, the reduced density  $\rho_{r}$ at time $t_{f}$ is rewritten as
\begin{equation}
    \rho_{r}(\mathbf{q}_{f},\tilde{\mathbf{q}}_{f},t_{f})=\int_{-\infty}^{\infty}d^3 \mathbf{q}_{1} d^3 \mathbf{q}_{2}\int_{\mathbf{q}_{1}}^{\mathbf{q}_{f}}\mathcal{D}\mathbf{q}^{+}\!\!\int_{\mathbf{q}_{2}}^{\tilde{\mathbf{q}}_{f}}\mathcal{D}\mathbf{q}^{-}\;e^{i\,S_{CG}[\mathbf{q}^{+},\mathbf{q}^{-}]}\, \rho_{e}(\mathbf{q}_{1},\mathbf{q}_{2},t_{i})\,,\label{E:pdrje}
\end{equation}
Here the coarse-grained action $S_{\text{CG}}$ is defined by
\begin{equation}
    S_{\text{CG}}[\mathbf{q}^{+},\mathbf{q}^{-}]=
S_{e}[\mathbf{q}^{+}]-S_{e}[\mathbf{q}^{-}]-i \ln
\mathcal{F}[\mathbf{j}^+_{\mathrm{T}},\mathbf{j}^-_{\mathrm{T}}] \,,
\label{CGact}
\end{equation}
in which $S_{e}[\mathbf{q}]$ is the action corresponding to the Lagrangian \eqref{lagrangian-e}. The resulting expression of $\mathcal{F}\left[\mathbf{j}^{+},\mathbf{j}^{-}\right]$ is then obtained in terms of real-time Green's functions of the vector potentials,
\begin{align}\label{E:kjkjkj}
    \mathcal{F}\left[\mathbf{j}^{+},\mathbf{j}^{-}\right]=\exp\biggl\{-\frac{1}{2\hbar^2}\int d^4x\!\!&\int\!d^4x'\Bigl[j^+_i(x;\mathbf{q}^+(t))\,\bigl<A^{+i}_{\mathrm{T}}(x)A^{+j}_{\mathrm{T}}(x')\bigr>\,j^+_j(x';\mathbf{q}^+(t'))\Bigr.\biggr.\notag\\
    &\qquad\qquad-j^+_i(x;\mathbf{q}^+(t))\,\bigl<A^{+i}_{\mathrm{T}}(x)A^{-j}_{\mathrm{T}}(x')\bigr>\,j^-_j(x';\mathbf{q}^-(t'))\notag\\
    &\qquad\qquad-j^-_i(x;\mathbf{q}^-(t))\,\bigl<A^{-i}_{\mathrm{T}}(x)A^{+j}_{\mathrm{T}}(x')\bigr>\,j^+_j(x';\mathbf{q}^+(t'))\notag\\
    &\biggl.\Bigl.+\;j^-_i(x;\mathbf{q}^-(t))\,\bigl<A^{-i}_{\mathrm{T}}(x)A^{-j}_{\mathrm{T}}(x')\bigr>\,j^-_j(x';\mathbf{q}^-(t'))\Bigr]\biggr\}\,,
\end{align}
The Green's functions in Eq.~\eqref{E:kjkjkj} are respectively
\begin{align*}
    \bigl<A^{+i}_{\mathrm{T}}(x)A^{+j}_{\mathrm{T}}(x')\bigr>&=\bigl<A_{\mathrm{T}}^i(x)A_{\mathrm{T}}^j(x')\bigr>\,\theta(t-t')+\bigl<A_{\mathrm{T}}^j(x')A_{\mathrm{T}}^i(x)\bigr>\,\theta(t'-t)\,,\\
    \bigl<A^{-i}_{\mathrm{T}}(x)A^{-j}_{\mathrm{T}}(x')\bigr>&=\bigl<A_{\mathrm{T}}^j(x')A_{\mathrm{T}}^i(x)\bigr>\,\theta(t-t')+\bigl<A_{\mathrm{T}}^i(x)A_{\mathrm{T}}^j(x')\bigr>\,\theta(t'-t)\,,\\
    \bigl<A^{+i}_{\mathrm{T}}(x)A^{-j}_{\mathrm{T}}(x')\bigr>&=\bigl<A_{\mathrm{T}}^j(x')A_{\mathrm{T}}^i(x)\bigr>\equiv\mathrm{Tr}\left\{\rho_{\mathbf{A}_{\mathrm{T}}}\,A_{\mathrm{T}}^j(x')A_{\mathrm{T}}^i(x)\right\}\,,\\
    \bigl<A^{-i}_{\mathrm{T}}(x)A^{+j}_{\mathrm{T}}(x')\bigr>&=\bigr<A_{\mathrm{T}}^i(x)A_{\mathrm{T}}^j(x')\bigr>\equiv\mathrm{Tr}\left\{\rho_{\mathbf{A}_{\mathrm{T}}}\,A_{\mathrm{T}}^i(x)A_{\mathrm{T}}^j(x')\right\}\,.
\end{align*}
We rewrite the imaginary part of $S_{\text{CG}}$ in terms of probability functional $\mathcal{P}[\mathbf{\xi}]$ of some stochastic noise $\mathbf{\xi}$,
\begin{equation} \label{p:xi}
    e^{-\operatorname{Im}\left\{S_{\text{CG}}[\mathbf{q}^{+},\mathbf{q}^{-}]\right\}}=\int\mathcal{D}\xi\;\mathcal{P}[\xi]\exp\left[-i\,e\int_{t_{i}}^{t_{f}}dt\;(
    \mathbf{q}^{+}-\mathbf{q}^{-})^{k}\left(\delta^{kl}\frac{d}{dt}-q^{l}\nabla^{k}\right)\xi^{l}\right]\,,
\end{equation}
where
\begin{equation}
    \mathcal{P}[\xi(t)]=\exp\left\{-\frac{\hbar}{2}\int_{-\infty}^{\infty}dt\int_{-\infty}^{\infty}dt'\;\left[\xi^i(t)\,G_{H}^{ij}{}^{-1}\left[{\bf q}(t),{\bf q}(t');t, t'\right]\,\xi^j(t')\right]\right\}\,
\end{equation}
and the Hadamard function $G_{H}^{ij}$ is defined in \eqref{Hadamrd}. From the derivations so far, we may interpret the stochastic noise $\xi$ as manifestation of quantum fluctuations of the environmental field. Hereafter it is convenient to introduce the stochastic coarse-grained effective action $S_{\xi}$ by
\begin{equation}
    S_{\mathbf{\xi}}[\mathbf{q}^{+},\mathbf{q}^{-}
    ;\xi]=\operatorname{Re}\left\{S_{\text{CG}}[\mathbf{q}^{+},\mathbf{q}^{-}]\right\}-e\int_{t_{i}}^{t_{f}}dt\;(
    \mathbf{q}^{+}-\mathbf{q}^{-})^{k}\left(\delta^{kl}\frac{d}{dt}-q^{l}\nabla^{k}\right)\xi^{l}
    \, .
\end{equation}
The reduced density \eqref{E:pdrje} then becomes
\begin{align}
     \rho_{r}(\mathbf{q}_{f},\tilde{\mathbf{q}}_{f},t_{f})&=\int_{-\infty}^{\infty}  d^3 \mathbf{q}_{1} \, d^3 \mathbf{q}_{2}\int^{q_{f}}_{q_{1}} \mathcal{D}\mathbf{q}^{+} \!\!\int_{q_{2}}^{\tilde{q}_{f}}\mathcal{D}\mathbf{q}^{-}\int\mathcal{D}\xi\;\mathcal{P}[\xi]\,e^{i\,S_{\xi}[\mathbf{q}^{+},\mathbf{q}^{-};\xi]}\rho_{e}(\mathbf{q}_{1},\mathbf{q}_{2},t_{i})\notag\\
        &=\int\mathcal{D}\xi\;\mathcal{P}[\xi]\,\rho_{r}(\mathbf{q}_{f},\tilde{\mathbf{q}}_{f},t_{f};\xi)\,,
\end{align}
in which $\rho_{r}(\mathbf{q}_{f},\tilde{\mathbf{q}}_{f},t_{f};\xi)$ is a reduced density of the particle under the influence of some realization of environmental stochastic noise $\xi$,
\begin{equation} \rho_{r}(\mathbf{q}_{f},\tilde{\mathbf{q}}_{f},t_{f};\xi)=\int_{-\infty}^{\infty}d^3\mathbf{q}_{1}\,d^3\mathbf{q}_{2}\int^{q_{f}}_{q_{1}}\mathcal{D}\mathbf{q}^{+}\!\!\int_{q_{2}}^{\tilde{q}_{f}}\mathcal{D}\mathbf{q}^{-}\;e^{i\,S_{\xi}[\mathbf{q}^{+},\mathbf{q}^{-};\xi]}\rho_{e}(\mathbf{q}_{1},\mathbf{q}_{2},t_{i})\,.
\end{equation}

\section{Derivation of Eq. \eqref{E:owiek}}\label{appenb}
We first observe that given a Hermitian operator~\cite{PF}
\begin{equation*}
    \mathcal{O}(x)=\sum_{\lambda}\Bigl\{a_{\lambda}^{\vphantom{\dagger}}\,h_{\lambda}^{\vphantom{*}}(x)+a_{\lambda}^{\dagger}\,h_{\lambda}(x)\Bigr\}\,,
\end{equation*}
for some well-behaved scalar function $h_{\lambda}(x)$, the expectation value $\langle\mathcal{O}^{\dagger}\mathcal{O}\rangle$ is always greater than or equal to zero for any state. The label $\lambda$ denotes some quantum number no matter discrete or continuous. The operators $a_{\lambda}$ and $a^{\dagger}_{\lambda}$ are annihilation and creation operators, satisfying $[a^{\vphantom{\dagger}}_{\lambda},a^{\dagger}_{\lambda'}]=\delta_{\lambda\lambda'}$ and zero otherwise. The consequence of positivity of $\langle\mathcal{O}^{\dagger}\mathcal{O}\rangle$,
\begin{align*}
    \langle\mathcal{O}^{\dagger}\mathcal{O}\rangle&=\sum_{\lambda\lambda'}\Bigl\{\langle a_{\lambda}a_{\lambda'}\rangle\,h_{\lambda}h_{\lambda'}+\langle a_{\lambda}a^{\dagger}_{\lambda'}\rangle\,h_{\lambda}h^*_{\lambda'}+\langle a_{\lambda}^{\dagger}a_{\lambda'}\rangle\,h^*_{\lambda}h_{\lambda'}+\langle a_{\lambda}^{\dagger}a_{\lambda'}^{\dagger}\rangle\,h^*_{\lambda}h^*_{\lambda'}\Bigr\}\\
    &=\sum_{\lambda\lambda'}\Bigl\{\langle a_{\lambda}a_{\lambda'}\rangle\,h_{\lambda}h_{\lambda'}+\langle a^{\dagger}_{\lambda'}a_{\lambda}\rangle\,h_{\lambda}h^*_{\lambda'}+\langle a_{\lambda}^{\dagger}a_{\lambda'}\rangle\,h^*_{\lambda}h_{\lambda'}+\langle a_{\lambda}^{\dagger}a_{\lambda'}^{\dagger}\rangle\,h^*_{\lambda}h^*_{\lambda'}\Bigr\}+\sum_{\lambda\lambda'}[a_{\lambda},a^{\dagger}_{\lambda'}]\,h_{\lambda}h^*_{\lambda'}\\
    &=\sum_{\lambda\lambda'}\Bigl\{\langle a_{\lambda}a_{\lambda'}\rangle\,h_{\lambda}h_{\lambda'}+\langle a^{\dagger}_{\lambda'}a_{\lambda}\rangle\,h_{\lambda}h^*_{\lambda'}+\langle a_{\lambda}^{\dagger}a_{\lambda'}\rangle\,h^*_{\lambda}h_{\lambda'}+\langle a_{\lambda}^{\dagger}a_{\lambda'}^{\dagger}\rangle\,h^*_{\lambda}h^*_{\lambda'}\Bigr\}+\sum_{\lambda}h_{\lambda}h^*_{\lambda}\geq0\,,
\end{align*}
implies
\begin{equation}
    \sum_{\lambda\lambda'}\Bigl\{\langle a_{\lambda}a_{\lambda'}\rangle\,h_{\lambda}h_{\lambda'}+\langle a^{\dagger}_{\lambda'}a_{\lambda}\rangle\,h_{\lambda}h^*_{\lambda'}+\langle a_{\lambda}^{\dagger}a_{\lambda'}\rangle\,h^*_{\lambda}h_{\lambda'}+\langle a_{\lambda}^{\dagger}a_{\lambda'}^{\dagger}\rangle\,h^*_{\lambda}h^*_{\lambda'}\Bigr\}\geq-\sum_{\lambda}h_{\lambda}h^*_{\lambda}\,.
\end{equation}
Therefore, in the context of velocity dispersion of a static charge in the dipole approximation, we have the scalar function $h_{\mathbf{k}}$
equivalently given by
\begin{equation*}
    h_{\mathbf{k}}=\int du\;\frac{e}{m}\sqrt{\frac{1}{2\, \omega}}\, f(u) \;e^{-i\omega u}
\end{equation*}
with $\omega=\lvert\mathbf{k}\rvert$, such that the velocity
dispersion takes the form
\begin{equation*}
    \langle\Delta v^2_i(t)\rangle_{\xi}=\frac{1}{2}\int\!\frac{d^3k}{(2\pi)^{\frac{3}{2}}}\!\int\!\frac{d^3k'}{(2\pi)^{\frac{3}{2}}}\Bigl\{\langle a_{\mathbf{k}}a_{\mathbf{k}'}\rangle\,h_{\mathbf{k}}h_{\mathbf{k}'}+\langle a_{\mathbf{k}}a^{\dagger}_{\mathbf{k}'}\rangle\,h_{\mathbf{k}}h^*_{\mathbf{k}'}+\langle a_{\mathbf{k}}^{\dagger}a_{\mathbf{k}'}\rangle\,h^*_{\mathbf{k}}h_{\mathbf{k}'}+\langle a_{\mathbf{k}}^{\dagger}a_{\mathbf{k}'}^{\dagger}\rangle\,h^*_{\mathbf{k}}h^*_{\mathbf{k}'}\Bigr\}+\text{c.c.}\,,
\end{equation*}
and the corresponding renormalized velocity dispersion
$\delta\langle\Delta v^2_i(t)\rangle$ is given by
\begin{align}
    \delta\langle\Delta v^2_i(t)\rangle_{\xi}&=\int\!\frac{d^3k}{(2\pi)^{\frac{3}{2}}}\!\int\!\frac{d^3k'}{(2\pi)^{\frac{3}{2}}}\Bigl\{\langle a_{\mathbf{k}}a_{\mathbf{k}'}\rangle\,h_{\mathbf{k}}h_{\mathbf{k}'}+\langle a^{\dagger}_{\mathbf{k}'}a_{\mathbf{k}}\rangle\,h_{\mathbf{k}}h^*_{\mathbf{k}'}+\langle a_{\mathbf{k}}^{\dagger}a_{\mathbf{k}'}\rangle\,h^*_{\mathbf{k}}h_{\mathbf{k}'}+\langle a_{\mathbf{k}}^{\dagger}a_{\mathbf{k}'}^{\dagger}\rangle\,h^*_{\mathbf{k}}h^*_{\mathbf{k}'}\Bigr\}\notag\\
    &\geq-\int\!\frac{d^3k}{(2\pi)^{3}}\;h_{\mathbf{k}}h^*_{\mathbf{k}}\, .\label{E:owkks}
\end{align}

\section{Evaluation of Eqs.~\eqref{E:sdfsk} and \eqref{E:dkjfhs}}\label{appendc}
Here we preform integration over $\tau$ and $\tau'$ variables in Eqs.~\eqref{E:sdfsk} and \eqref{E:dkjfhs} which respectively give rise to the expressions for the stationary component of the change in velocity fluctuations
\begin{equation}\label{E:stw}
    \delta\langle\Delta v_i^2(t)\rangle_{\xi,\,\text{st}}=Z^2\frac{e^2}{m^2}A(d\Omega_{s})\,\eta^{2}\int_{\Xi-\Delta/2}^{\Xi+\Delta/2}d\omega\;\frac{\omega^3}{4}\left\{L_1(\omega)+L_2(\omega)+L_3(\omega)+L_4(\omega)\right\}\,,
\end{equation}
where
\begin{align*}
    L_1(\omega)&=\frac{1}{2\Gamma(\Omega+i\Gamma)(\omega-\Omega-i\Gamma)}\\
    &\qquad\times\biggl[-i\Bigl(\Omega+i\Gamma\Bigr)\Bigl(1+e^{-2\Gamma t}-e^{-\Gamma t+i\omega t-i\Omega t}-e^{-\Gamma t-i\omega t+i\Omega t}\Bigr)\biggr.\\
    &\qquad\qquad\qquad\biggl.+\;e^{+2i\delta}\,\Gamma\Bigl(1-e^{-\Gamma t-i\omega t+i\Omega t}-e^{-\Gamma t+i\omega t+i\Omega t}+e^{-2\Gamma t+i2\Omega t}\Bigr)\biggr]\,,\\
    L_2(\omega)&=\frac{1}{2\Gamma(\Omega-i\Gamma)(\omega-\Omega+i\Gamma)}\\
    &\qquad\times\biggl[+i\Bigl(\Omega-i\Gamma\Bigr)\Bigl(1+e^{-2\Gamma t}-e^{-\Gamma t+i\omega t-i\Omega t}-e^{-\Gamma t-i\omega t+i\Omega t}\Bigr)\biggr.\\
    &\qquad\qquad\qquad\biggl.+\;e^{-2i\delta}\,\Gamma\Bigl(1-e^{-\Gamma t+i\omega t-i\Omega t}-e^{-\Gamma t-i\omega t-i\Omega t}+e^{-2\Gamma t-i2\Omega t}\Bigr)\biggr]\,,\\
    L_3(\omega)&=\frac{1}{2\Gamma(\Omega+i\Gamma)(\omega+\Omega+i\Gamma)}\\
    &\qquad\times\biggl[+i\Bigl(\Omega+i\Gamma\Bigr)\Bigl(1+e^{-2\Gamma t}-e^{-\Gamma t-i\omega t-i\Omega t}-e^{-\Gamma t+i\omega t+i\Omega t}\Bigr)\biggr.\\
    &\qquad\qquad\qquad\biggl.-\;e^{+2i\delta}\,\Gamma\Bigl(1-e^{-\Gamma t-i\omega t+i\Omega t}-e^{-\Gamma t+i\omega t+i\Omega t}+e^{-2\Gamma t+i2\Omega t}\Bigr)\biggr]\,,\\
    L_4(\omega)&=\frac{1}{2\Gamma(\Omega-i\Gamma)(\omega+\Omega-i\Gamma)}\\
    &\qquad\times\biggl[-i\Bigl(\Omega-i\Gamma\Bigr)\Bigl(1+e^{-2\Gamma t}-e^{-\Gamma t-i\omega t-i\Omega t}-e^{-\Gamma t+i\omega t+i\Omega t}\Bigr)\biggr.\\
    &\qquad\qquad\qquad\biggl.-\;e^{-2i\delta}\,\Gamma\Bigl(1-e^{-\Gamma t+i\omega t-i\Omega t}-e^{-\Gamma t-i\omega t-i\Omega t}+e^{-2\Gamma t-i2\Omega t}\Bigr)\biggr]\,,
\end{align*}
and for the nonstationary component
\begin{equation}\label{E:nsw}
    \delta\langle\Delta v_i^2(t)\rangle_{\xi,\,\text{ns}}=Z^2\frac{e^2}{m^2}A(d\Omega_{s})\,\mu\eta\int_{\Xi-\Delta/2}^{\Xi+\Delta/2}d\omega\;\frac{\omega^3}{4}\left\{J_1(\omega)+J_2(\omega)+J_3(\omega)\right\}\,,
\end{equation}
where
\begin{align*}
    J_{1}(\omega)&=e^{-2i\delta}\left[-\frac{e^{-2i\omega t+i\theta}}{2(\omega+i\Gamma-\Omega)^2}+\frac{e^{-\Gamma t-i\omega t-i\Omega t+i\theta}}{(\omega+i\Gamma-\Omega)^2}-\frac{e^{-2\Gamma t-2i\Omega t+i\theta}}{2(\omega+i\Gamma-\Omega)^2}\right]\\
    &\qquad\qquad+e^{2i\delta}\left[-\frac{e^{2i\omega t-i\theta}}{2(\omega-i\Gamma-\Omega)^2}+\frac{e^{-\Gamma t+i\omega t+i\Omega t-i\theta}}{(\omega-i\Gamma-\Omega)^2}-\frac{e^{-2\Gamma t+2i\Omega t-i\theta}}{2(\omega-i\Gamma-\Omega)^2}\right]\,,\\
    J_2(\omega)&=\left[-\frac{e^{-2\Gamma t-i\theta}}{(\omega-i\Gamma)^2-\Omega^2}-\frac{e^{2i\omega t-i\theta}}{(\omega-i\Gamma)^2-\Omega^2}+\frac{e^{-\Gamma t+i\omega t-i\Omega t-i\theta}}{(\omega-i\Gamma)^2-\Omega^2}+\frac{e^{-\Gamma t+i\omega t+i\Omega t-i\theta}}{(\omega-i\Gamma)^2-\Omega^2}\right.\\
    &\qquad\left.-\frac{e^{-2\Gamma t+i\theta}}{(\omega+i\Gamma)^2-\Omega^2}-\frac{e^{-2i\omega t+i\theta}}{(\omega+i\Gamma)^2-\Omega^2}+\frac{e^{-\Gamma t-i\omega t-i\Omega t+i\theta}}{(\omega+i\Gamma)^2-\Omega^2}+\frac{e^{-\Gamma t-i\omega t+i\Omega t+i\theta}}{(\omega+i\Gamma)^2-\Omega^2}\right]\,,\\
    J_3(\omega)&=e^{-2i\delta}\left[-\frac{e^{2i\omega t-i\theta}}{2(\omega-i\Gamma+\Omega)^2}+\frac{e^{-\Gamma t+i\omega t-i\Omega t-i\theta}}{(\omega-i\Gamma+\Omega)^2}-\frac{e^{-2\Gamma t-2i\Omega t-i\theta}}{2(\omega-i\Gamma+\Omega)^2}\right]\\
    &\qquad\qquad+e^{2i\delta}\left[-\frac{e^{-2i\omega t+i\theta}}{2(\omega+i\Gamma+\Omega)^2}+\frac{e^{-\Gamma t-i\omega t+i\Omega t+i\theta}}{(\omega+i\Gamma+\Omega)^2}-\frac{e^{-2\Gamma t+2i\Omega t+i\theta}}{2(\omega+i\Gamma+\Omega)^2}\right]\,.
\end{align*}


\begin{thebibliography}{99}

\bibitem{EP}
H. Epstein, V. Glaser, and A. Jaffe, Nuovo Cimento {\bf 36}, 1016 (1965).

\bibitem{MO}
M. S. Morris and K. S. Thorne, Am. J. Phys. {\bf 56}, 395 (1988);  M. S. Morris, K. S. Thorne, and U. Yurtsever, Phys. Rev. Lett. {\bf 61}, 1446 (1988).

\bibitem{AL}
M. Alcubierre, Class. Quantum Grav. {\bf 11}, L73 (1994).

\bibitem{FOR}
L. H. Ford and T. A. Roman, Phys. Rev. D {\bf 41}, 3662 (1990) ; L. H. Ford and T. A. Roman, Phys. Rev. D {\bf 46}, 1328 (1992).

\bibitem{FO1}
L. H. Ford, Phys. Rev. D {\bf 43}, 3972 (1991).

\bibitem{PF}
M. J. Pfenning and L. H. Ford, Phys. Rev. D {\bf 57}, 3489 (1998) ; L. H. Ford, M. J. Pfenning, and T. A. Roman, Phys. Rev. D {\bf 57}, 4839 (1998).

\bibitem{FE}
C. J. Fewster and S. P. Eveson, Phys. Rev. D {\bf 58}, 084010 (1998).

\bibitem{FO}
L. H. Ford, Proc. R. Soc. London A {\bf 346}, 227 (1978).

\bibitem{DA}
P. C. W. Davies, Phys. Lett. B {\bf 113}, 393 (1982).

\bibitem{DAP}
E. W. Davis and H. E. Puthoff, {\it Experimental Concepts for Generating Negative Energy in the Laboratory},  AIP Conf. Proc. {\bf 813}, 1362 (2006).


\bibitem{BO}
D. Boyanovsky, H. J. de Vega, and R. Holman, Phys. Rev. D {\bf 49}, 2769 (1994); D. Boyanovsky, H. J. de Vega, R. Holman, D.-S. Lee, and A. Singh, Phys. Rev. D {\bf 51}, 4419 (1995); D. Boyanovsky, M. D'Attanasio, H. J. de Vega, R. Holman, and D.-S. Lee,  Phys. Rev. D {\bf 52}, 6805 (1995).

\bibitem{DAO}
P. C. W. Davies and A. C. Ottewill, Phys. Rev. D {\bf 65}, 104014 (2002).

\bibitem{MA}
P. Marecki, Phys. Rev. A {\bf 66}, 053801 (2002); P. Marecki and N. Szpak, Ann. Phys. (Leipzig) {\bf 14}, 428 (2005).

\bibitem{FOG}
L. H. Ford, P. G. Grove, and A. C. Ottewill, Phys. Rev. D {\bf 46}, 4566 (1992).

\bibitem{HF}
J.-T. Hsiang and L. H. Ford, Phys. Rev. D {\bf 78}, 065012 (2008).

\bibitem{FV}
R. P. Feynman and F. Vernon, Ann. Phys. (N.Y.) {\bf 24}, 118 (1963).

\bibitem{SK}
J. Schwinger, J. Math. Phys. {\bf 2}, 407 (1961); L.V. Keldysh, Sov. Phys. JETP {\bf 20}, 1018 (1965).

\bibitem{GR1}
H. Grabert, P. Schramm, and G. L. Ingold, Phys. Rep. {\bf 168}, 115 (1988).

\bibitem{BA}
P. M. V. B. Barone and A. O. Caldeira, Phys. Rev. A {\bf 43}, 57 (1991).



\bibitem{WU}
C.-H. Wu and D.-S. Lee,  Phys. Rev. D {\bf 71}, 125005 (2005).

\bibitem{JO}
P. R. Johnson and B. L. Hu, Phys. Rev D {\bf 65}, 065015 (2002).

\bibitem{HUP}
B. L. Hu and D. Pavon, Phys. Lett. B {\bf 180}, 329 (1986); B. L. Hu and H. E. Kandrup, Phys. Rev. D {\bf 35}, 1776 (1987); H. E. Kandrup, Phys. Rev. D {\bf 37}, 3505 (1988); B. L. Hu and A. Matacz, Phys. Rev. D {\bf 49}, 6612 (1994).

\bibitem{JT}
J.-T. Hsiang and D.-S. Lee, Phys. Rev. D {\bf 73} 065022 (2006).

\bibitem{HLW}
J.-T. Hsiang, D.-S. Lee, and C.-H. Wu, J. Korean  Phys. Soc. {\bf 49}, 742 (2006).

\bibitem{HWL}
J.-T. Hsiang, T.-H. Wu, and D.-S. Lee, Phys. Rev. D {\bf 77} 105021 (2008).

\bibitem{ITZ}
C. Itzykson and J.-B. Zuber, {\it Quantum Field Theory} (McGraw-Hill, 1980).

\bibitem{SAK}
J. J. Sakurai, {\it Advanced Quantum Mechanics} (Addison-Wesley, 1967).

\bibitem{YF}
H. Yu and L. H. Ford, Phys. Rev. D {\bf 70}, 065009 (2004).

\bibitem{HZ}
B. L. Hu and Y.-H. Zhang, Int. J. Mod. Phys. A {\bf 10}, 4537 (1995).

\bibitem{ABV}
S. M. Alamoudi, D. Boyanovsky, H. J. de Vega, and R. Holman, Phys.
Rev. D {\bf 59}, 025003 (1998).

\bibitem{JA}
J. D. Jackson, {\it Classical Electrodynamics, 3rd Edition}, (Wiley, 1998).

\bibitem{Hak}
V. Hakim and V. Ambegaokar, Phys. Rev. A {\bf 32}, 423 (1985).

\bibitem{BRE}
H.-P. Breuer and F. Petruccione, {\it The Theory of Open Quantum Systems} (Oxiford University Press, 2002).



\bibitem{MN}
A. B. Matsko, I. Novikova, G. R. Welch, D. Budker, D. F. Kimball, and S. M. Rochester, Phys. Rev. A {\bf 66}, 043815 (2002); E. E. Mikhailov and I. Novikova, Optics Letters, {\bf 33}, 1213 (2008).

\bibitem{CAL}
A. O. Caldeira and A. J. Leggett, Physica A {\bf 121}, 587 (1983); Ann. Phys. (N.Y.) {\bf 149}, 374 (1983); A. J. Leggett et al., Rev. Mod. Phys. {\bf 59}, 1 (1987).

\bibitem{HPZ}
B. L. Hu, J. P. Paz, and Y.-H. Zhang, Phys. Rev. D {\bf 45}, 2843
(1992).

\bibitem{WHL}
T.-H. Wu, J.-T. Hsiang, and D.-S. Lee, Found. Phys.,{\bf 41}, 77 (2011).

\end{thebibliography}
\end{document}